\documentclass[sigconf]{acmart}
\usepackage{amsmath,amsfonts}
\usepackage{algorithmic}
\usepackage{graphicx}
\usepackage{subfig}
\usepackage{textcomp}
\usepackage{xcolor}
\usepackage{gensymb}

\usepackage{soul, color}
\soulregister{\cite}7 %
\soulregister{\ref}7 %

\AtBeginDocument{%
  \providecommand\BibTeX{{%
    \normalfont B\kern-0.5em{\scshape i\kern-0.25em b}\kern-0.8em\TeX}}}

 \renewcommand\footnotetextcopyrightpermission[1]{}

\setcopyright{acmcopyright} 
\copyrightyear{2018}
\acmYear{2018}
\acmDOI{XXXXXXX.XXXXXXX}

\acmConference[IPSN '24]{The 23rd ACM/IEEE International Conference on Information Processing in Sensor Networks}{May 13--16,
  2024}{HONG KONG, CHINA}
\acmPrice{15.00}
\acmISBN{978-1-4503-XXXX-X/18/06}

\begin{document}

\title{NNCTC: Physical Layer Cross-Technology Communication via Neural Networks}
\thanks{\dag Co-primary student authors. * Corresponding author}

\author{Haoyu Wang}
\authornotemark[2]
\email{haoy_w@njfu.edu.cn}
\affiliation{%
  \institution{Nanjing Forestry University}
  \city{Nanjing}
  \country{China}
}

\author{Jiazhao Wang}
\authornotemark[2]
\email{jiazhao_wang@mymail.sutd.edu.sg}
\affiliation{%
  \institution{Singapore University of Technology and Design}
  \country{Singapore}
}

\author{Demin Gao}
\email{dmgao@njfu.edu.cn}
\affiliation{%
  \institution{Nanjing Forestry University}
  \city{Nanjing}
  \country{China}
}

\author{Wenchao Jiang}
\authornotemark[1]
\email{wenchao_jiang@sutd.edu.sg}
\affiliation{%
  \institution{Singapore University of Technology and Design}
  \country{Singapore}
}

\begin{abstract}
Cross-technology communication~(CTC) enables seamless interactions between diverse wireless technologies. Most existing work is based on reversing the transmission path to identify the appropriate payload to generate the waveform that the target devices can recognize. However, this method suffers from many limitations, including dependency on specific technologies and the necessity for intricate algorithms to mitigate distortion. In this work, we present NNCTC, a Neural-Network-based Cross-Technology Communication framework inspired by the adaptability of trainable neural models in wireless communications. By converting signal processing components within the CTC pipeline into neural models, the NNCTC is designed for end-to-end training without requiring labeled data. This enables the NNCTC system to autonomously derive the optimal CTC payload, which significantly eases the development complexity and showcases the scalability potential for various CTC links. Particularly, we construct a CTC system from Wi-Fi to ZigBee. The NNCTC system outperforms the well-recognized WEBee and WIDE design in error performance, achieving an average packet reception rate~(PRR) of 92.3\% and an average symbol error rate~(SER) as low as 1.3\%.

\end{abstract}

\maketitle

\section{Introduction} 
The Internet of Things (IoT) is showing strong momentum with a soaring number of IoT devices estimated to exceed 43 billion by 2023~\cite{rahmani2023next}. The huge number of IoT devices has led to ubiquitous interference, especially on the free spectrum (such as the 2.4GHz ISM frequency band).  But at the same time, it also brings opportunities for collaboration between different wireless communication technologies, represented by cross-technology communication (CTC)~\cite{li2017webee}. CTC achieves direct communication between different wireless communication technologies in the physical layer (PHY). The latest work aims to establish more robust CTC connections and, based on this foundation, implement a range of new applications, such as~\cite{yin2018explicit, gao2022spectrum, gao2023time}.

Despite the effort and clever designs to promote CTC connectivities~\cite{kim2015freebee, li2017webee}, existing CTC schemes encounter two fundamental challenges that hinder their popularity:
\begin{itemize}
    \item \textbf{Technology-specific emulation:} Existing CTC schemes are usually designed in a case-by-case manner for a certain pair of wireless technologies. Though interesting characteristics in the PHY layer are discovered and employed for communication, such an approach is hard to extend to general cases. In addition, existing CTC schemes tend to adopt a white-box manner in signal emulation, which requires a full understanding of the whole signal processing procedure and hardware implementation details. It not only has a steep learning curve but also leads to the potential lock-in of a specific software stack or supplier's equipment.

    \item \textbf{Hand-crafted parameters:} Choosing the optimal parameters is another challenge for existing CTC schemes. First, high-end wireless technology usually has a lot of parameters in its physical layer configuration. For example, an 802.11g/n transmitter needs to carefully choose its symbols and subcarriers to achieve optimal CTC communication to a ZigBee receiver \cite{li2017webee}. Second, even if the optimal setting is found through some optimization methods, it's hard for the CTC scheme to adapt to the complex communication environment, which restricts its performance in harsh wireless environments. 

\end{itemize}

We are inspired by the recent advances in applying AI in physical layer communication \cite{hoydis2022sionna, o2017introduction, he2019model}. Thanks to the powerful approximation capability of the neural network, an AI model can approximate signal processing blocks, complex channel models, or work as a whole end-to-end communication system \cite{mu2019end, hoydis2022sionna}. When AI meets CTC, we find the keys to the problems that have long troubled CTC.

In this work, we propose Neural-Network-based Cross-Technology Communication~(NNCTC), a general framework to construct CTC with the help of neural networks. We reinvestigate classic CTC strategies to demonstrate how to fit them into such a framework and bring their scalability, flexibility, and learning ability to the next level.  We believe such a novel framework will shed light on solutions to CTC design bottlenecks and advance the broader adoption of such a technology. Without loss of generality, we take the physical layer CTC from WiFi to ZigBee as an example to introduce in detail how NNCTC improves physical layer CTC with an end-to-end neural network. For a better demonstration of NNCTC, we consider the OFDM-based Wi-Fi schemes, those applying IEEE 802.11a/g/n/ac, which are more complex than CCK-based IEEE~802.11b. It's worth pointing out that NNCTC does not have to be limited to certain WiFi transmission schemes, as we design the NNCTC based on a generalized and recognized idea to emulate the time-domain waveforms.

In summary, the main contributions of this paper are summarized as follows:
\begin{itemize}
    \item Conceptually, this paper proposes the general framework of NNCTC aiming at easing CTC emulation and parameter setting. 

    \item Technologically, this paper demonstrates for the first time the usability and advantages of lightweight and interpretable NN in CTC.
 
    \item Experimentally, we implement NNCTC on USRP and show that it is applicable to multiple modulation schemes and outperforms handcrafted CTC technologies.

\end{itemize}

The remainder of this paper is organized as follows. In Section~\ref{sec: motivation} we introduce the motivation for writing this paper, including the existing problems, opportunities, and challenges. In Section~\ref{sec: overview} we give a macro overview of the NNCTC framework. In Section~\ref{sec:section_Emulation_qam}, we introduce in detail how to employ neural networks to implement the physical layer CTC and specifically focus on NN-based QAM emulation. Section~\ref{section:post-qam-channel-coding} presents the other two emulations required (Post-QAM emulation and Channel Coding emulation). In Section~\ref{sec: evaluation} we conduct extensive evaluation experiments. In Section~\ref{sec: DISCUSSION}, we have a related discussion. In Section~\ref{sec: related_work} we list a series of related works. Finally, Section~\ref{sec: conclusion} summarizes the research conclusions of this paper.

\section{Motivation}\label{sec: motivation}
\subsection{Restatement of Existing Issues}
CTC is to achieve direct communication between different protocols by emulating the time-domain waveforms that the target devices can detect and recognize. CTC can ease the requirement of centralized IoT gateways for different wireless techniques to communicate. The representative work in the field of CTC~\cite{li2017webee} proposes a strategy of using physical layer emulation to realize CTC communication from WiFi to ZigBee, where the transmission process of Wi-Fi OFDM modulator is reversed so that we can derive the payload that can generate a similar waveform(Referred to as signal simulation technology). The same idea of waveform emulation is supposed to be extended to other pairs of CTC links. However, when we intend to apply it to a new target wireless technique, the mitigation is not intuitive and direct. We need to re-design the emulation process, including segmenting the target signals, selecting the effective QAM symbols, and so on. To ease the manual progress as in WEBee, the latest CTC work~\cite{liao2023xituxi} applying the machine learning techniques by modeling the CTC process as a neural machine translation task, which requires a \textit{CTC corpora} from different wireless specifications and applies the emerging \textit{transformer} architecture with massive trainable parameters. Although it eases the development complexity of mitigation, it relies on a huge amount of training data and a large-scale neural network model.

In general, most of the current CTC research work is static, which means the parameter optimization procedure is technique-specific, while some of the latest work focus on modeling the CTC process through black-box machine learning models. With the progress of the neural network applied in the physical layer, we hope to find a lightweight and intelligent neural network-based solution to CTC waveform emulation.

\subsection{Opportunities}
\textbf{Neural networks can provide the learning ability for wireless communication systems.} Recently, the neural network~(NN) models have been widely adopted in the physical layer design for wireless communication~\cite{wen2018deep, wu2018development, lyu2018performance}. The neural network models are used to either replace certain blocks in the transceiver~\cite{wang2023demo}, or replace the entire receiver~\cite{zhao2021deep}, or even the whole transceiver~\cite{o2017introduction}. These models are trained to achieve different tasks, including modulation/demodulation, decoding, and so on. The great potential to learn for wireless communication tasks inspires us to utilize neural networks to construct the CTC pipeline.

\textbf{Waveform Emulation can perform effectively.} Since the birth of the physical-level CTC proposed in WEBee, most of the research work follows a similar idea to manipulate the payload to emulate the waveforms that can be detected and recognized by the target devices. The objective of the waveform emulation can be modeled by minimizing the error between the emulated waveform and the desired waveform.

Inspired by the learning ability of the NN-based physical layer design and the applicable design goal based on waveform emulation, we intend to construct an NN-based CTC emulation model, where the model is trained in an end-to-end way to reconstruct the targeting waveform.

\subsection{Challenges}
To implement lightweight neural networks for CTC emulation tasks, we are supposed to carefully design the model structure and training methodology, which is challenging.

The first challenge is to represent the processing blocks in the conventional CTC process with neural network models. To derive the desired payload, CTC systems are supposed to reverse the transmission process and find potential payloads that can generate the desired waveform. In~\cite{liao2023xituxi}, researchers implement the CTC processes in a black-box way by constructing them with large transformer models. Although such design outperformed conventional CTC design in accuracy and generalization capability, the overall network structure suffers high computation complexity conducted on powerful platforms~(NVIDIA RTX3080Ti GPU as declared in the paper). To address this challenge, we apply the model-driven approach to design the neural network models. More specifically, we start with the mathematical model of the signal processing blocks and discover the suitable neural network models for each block.

The other challenge is the training methodology for NN-based CTC. In \cite{liao2023xituxi}, the authors apply the idea from Large Language Model~(LLM) to train the transformer-based architecture, of which the training procedure requires a huge amount of both time and data. To ease the training procedure, we also follow the intuitive goal of waveform emulation in our design. More specifically, we implement an end-to-end structure by stacking the NN-based processing blocks together following the conventional CTC pipeline. The training goal is to make the output of the NN-based CTC as approximate as possible to the input waveform samples. Meanwhile, to reduce the training complexity, parameters from some NN-based blocks are fixed, and some constraints are also applied to the NN-based blocks to make the training convergent and parameters meaningful.

\section{Overview}\label{sec: overview}
\begin{figure}[!tbp]
    \centering
    \includegraphics[width=1.00\linewidth]{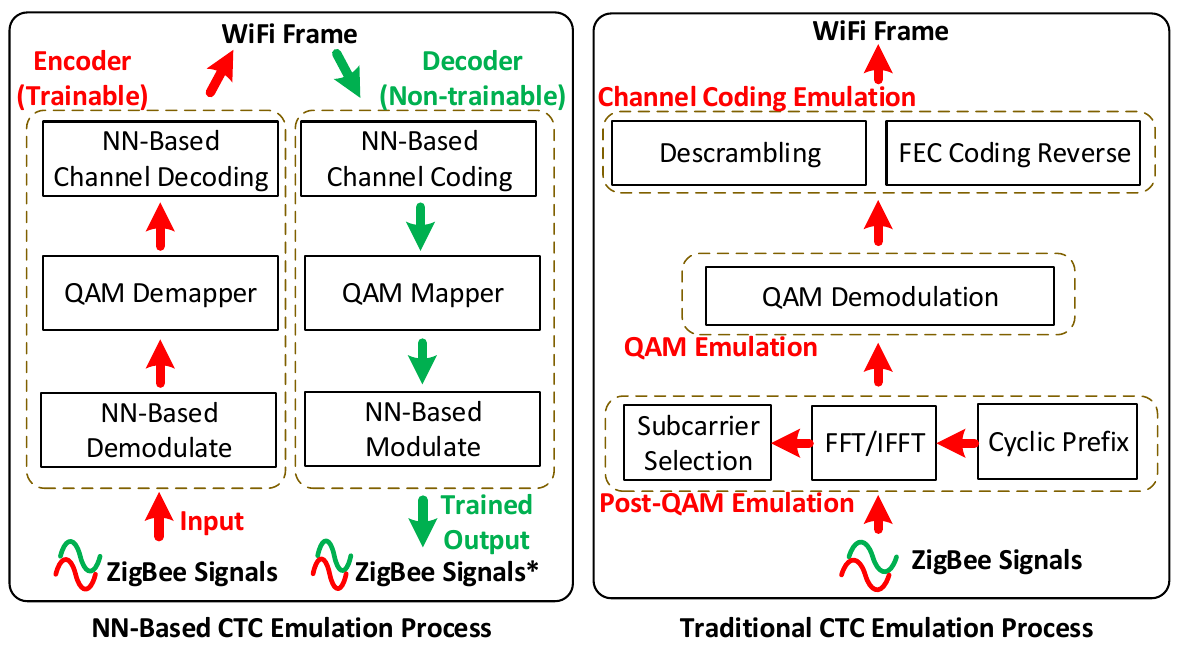}
    \caption{Workflow of NNCTC and conventional CTC}
    \label{fig:NNCTC_Overview}
    \vspace{-5mm}
\end{figure}

We take the physical layer CTC from the Wi-Fi system to the ZigBee system as a case study. More specifically, the payload derived from NNCTC can be processed by the OFDM-based modulator in Wi-Fi systems to generate signals that the O-QPSK-based ZigBee receiver can process. 

\textbf{Conventional CTC.}
The overview of the typical workflow of NNCTC, as well as conventional CTC, is depicted in Fig.~\ref{fig:NNCTC_Overview}. The image on the right of Fig.~\ref{fig:NNCTC_Overview} shows how conventional CTC emulates waveforms. It starts with ZigBee signals and ends with a WiFi Frame. QAM emulation mainly solves the problem of how to select QAM symbols after FFT. Post-QAM emulation tackles extra concerns post a physical signal change, like CP and pilot subcarrier choice.

\textbf{NNCTC.}
In contrast, the right side of Fig.~\ref{fig:NNCTC_Overview} depicts our emulation process based on Neural Network (NN). We have devised a bidirectional emulation, commencing with standard ZigBee signals, and "ZigBee Signals*" representing the output of the end-to-end model. Training of the encoder is accomplished by minimizing the gap between "ZigBee Signals*" and "ZigBee Signals," ultimately yielding the desired WiFi Frame.
We will describe how NNCTC implements an emulation path similar to traditional CTC. In Sec.~\ref{sec:section_Emulation_qam}, we will elaborate on how NNCTC implements QAM emulation. In Sec.~\ref{section:post-qam-channel-coding}, we will discuss how NNCTC achieves Post-QAM emulation and channel coding emulation.

\section{Realization of Physical Layer CTC via  Neural Network}\label{sec:section_Emulation_qam}

In this section, we will introduce how to implement physical layer CTC using neural networks. We focus on modeling the OFDM modulation/demodulation and QAM Mapping/Demapping processes through neural networks. We start with the mathematical model and transform it into the neural network structure. Then, we will demonstrate how CTC emulation from Wi-Fi to ZigBee can be achieved using the neural networks mentioned above.

\subsection{Mathematical Foundations of Transitioning from the Physical Layer to Neural Networks}\label{sec:Mathematical}
Before transitioning the physical layers of WiFi and ZigBee to neural networks, we need to establish some mathematical theoretical foundations. Earlier, it was mentioned that there are two approaches to using neural networks to implement the physical layer. The first approach is data-driven and requires large datasets to be fed into complex network models, which doesn't align with the lightweight requirements of IoT communication. The other approach is model-driven, where lightweight neural networks are constructed by uncovering the inherent connections between the mathematical foundations of communication and neural networks.

\paragraph{Mathematical Basis of NNCTC}
In wireless communication, we can model a wide range of modulation schemes with Signal Space Analysis, including single carrier amplitude/phase modulation as well as the Orthogonal Frequency Division Multiplexing~(OFDM) scheme. Based on this method, we can regard the modulation process (symbol to signal) as a linear combination of the symbol $s_{i}$ and the basis functions corresponding to each dimension \cite{wang2023demo,proakis2008digital,goldsmith2005wireless}, which can be expressed as:
\begin{equation}
    S_i(t) = \sum_{j=1}^{N}s_{ij}\phi_j(t)
    \label{eq:signal_space}
\end{equation}

The $s_{ij}$ in the formula represents the $j$-th dimension of the $s_{i}$, and $S_{i}(t)$ represents the modulated time-domain signal. The most important $\phi_j(t)$ in the formula represents the $j$-th function in the basis function set $\{\phi(t)\}^{N}$.

\paragraph{Neural Network Driven by Mathematical Model}
To build a neural network based on a mathematical model, we need to fit the mathematical model into the neural network. We first convert the continuous form in Equation~\ref{eq:signal_space} into discrete-time form, which is given as
\begin{equation}
\centering
   S_i[n] = \sum_{j=1}^{N}s_{ij}\phi_j[n]
\label{eq:signal_space_Discrete}
\end{equation}
$S_i[n]$ and $\phi_j[n]$ represent the samples of the original signal $S_i(t)$ and basis functions $\phi_j(t)$. 
We further expand our analysis to a general case where the symbols and the basis functions are complex-valued. By decomposing the complex-valued signal samples into real and imaginary parts, we can get
\begin{equation}
    \centering
    \begin{aligned}
        & S_I[n]+jS_Q[n]=\mathrm{Re}\{S_i[n]\}+j\mathrm{Im}\{S_i[n]\} \\
        & = \sum_{j=1}^N [\mathrm{Re}\{s_{ij}\}+j\mathrm{Im}\{s_{ij}\}][\mathrm{Re}\{\phi_{j}[n]\}+j\mathrm{Im}\{\phi_{j}[n]\}] \\
        & = \sum_{j=1}^N[\mathrm{Re}\{s_{ij}\}\mathrm{Re}\{\phi_{j}[n]\}-\mathrm{Im}\{s_{ij}\}\mathrm{Im}\{\phi_{j}[n]\}] \\
        & + \sum_{j=1}^N[\mathrm{Re}\{s_{ij}\}\mathrm{Im}\{\phi_{j}[n]\}+\mathrm{Im}\{s_{ij}\}\mathrm{Re}\{\phi_{j}[n]\}]
    \end{aligned}
    \label{eq:signal_space_complex}
\end{equation}
\begin{figure}[!tbp]
    \centering
    \includegraphics[width=1.0\linewidth]{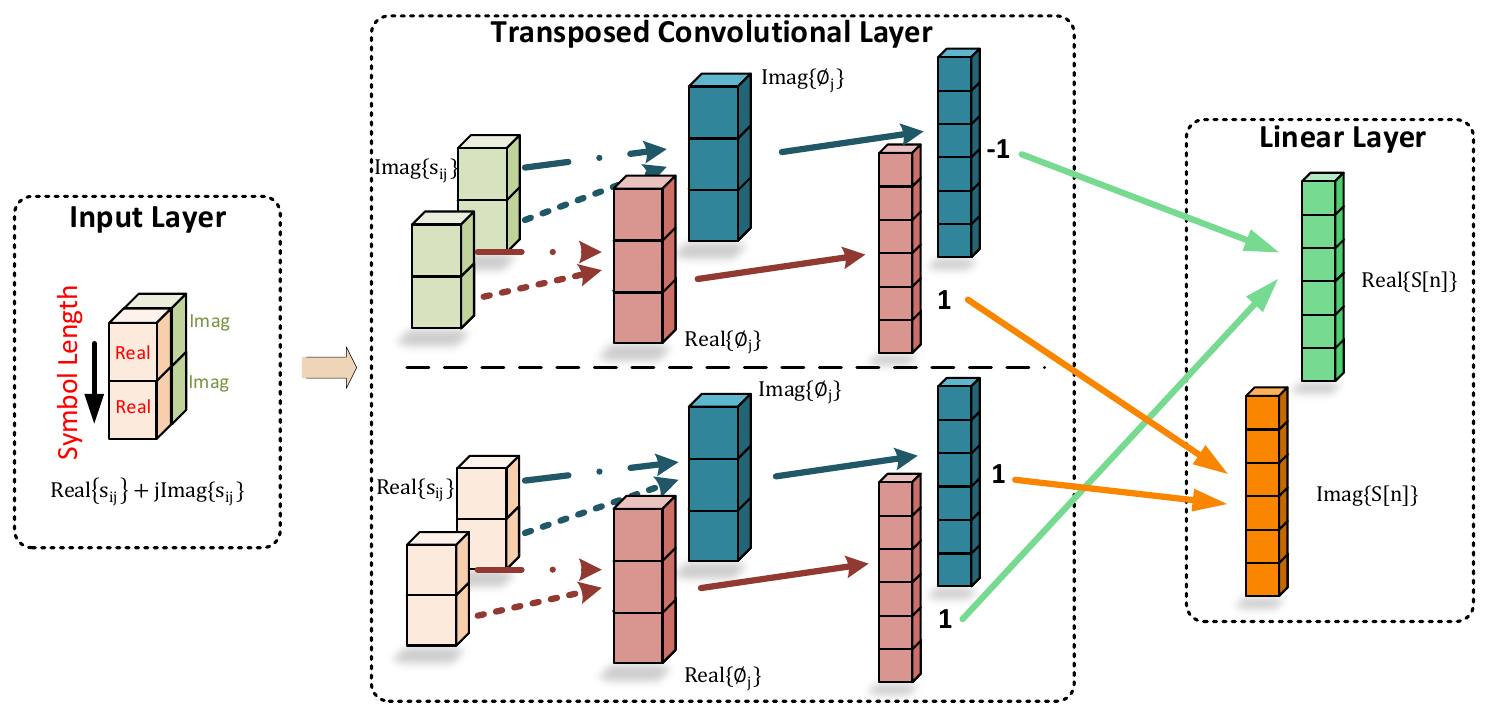}
    \caption{Configuration examples from mathematical foundations to neural networks}
    \label{fig:math-to-nn}
    \vspace{-3mm}
\end{figure}
To accommodate the modulation process into neural network layers, we apply the transposed convolutional layer and the linear layer. As shown in Fig.~\ref{fig:math-to-nn}, the real and imaginary parts are fed into the transposed convolutional layer. The kernel parameters of the transposed convolutional layer are set based on the basis functions, and the output from the transposed convolutional layer is the component of signal samples. Then, these components are combined through a fully connected layer to generate the modulated signals. 

\subsection{The Transformation of the WiFi Physical Layer into Neural Networks}

\begin{figure}[!tbp]
    \centering
    \includegraphics[width=1.00\linewidth]{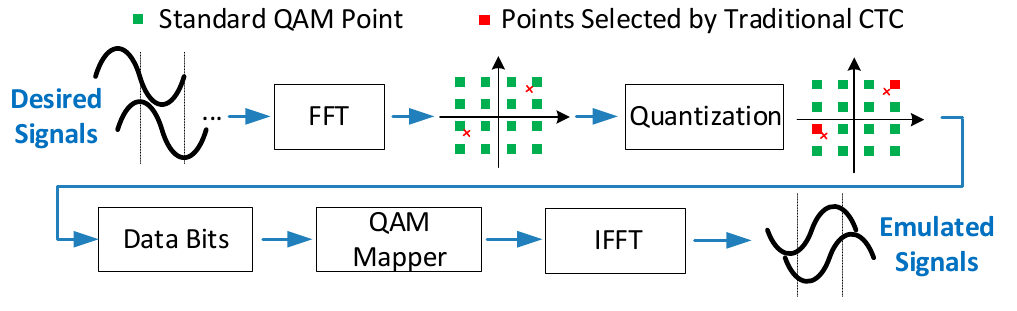}
    \caption{Conventional CTC QAM Emulation Process}
    \label{fig:qam-emulation-tran}
    \vspace{-3mm}
\end{figure}

Following the procedures in Fig.~\ref{fig:qam-emulation-tran}, we will convert the processing blocks into neural networks, including DFT/IDFT, QAM mapper, and QAM demapper. 

\paragraph{Convert DFT and IDFT processes to neural networks} DFT/IDFT processes share similar formulas as
\begin{equation}
    \centering
    \begin{aligned}
        &DFT:\quad X[n] = \sum_{i=1}^{N-1}S[n]e^{-\frac{j2\pi n i}{N}}, \quad 0\leq n \leq N-1 \\
        &IDFT: \quad S[n] = \sum_{i=1}^{N-1}X[n]e^{\frac{j2\pi n i}{N}}, \quad 0\leq n \leq N-1
    \end{aligned}
\end{equation} where $X[n]$ and $S[n]$ represent the frequency-domain components and the time-domain signals, respectively. $e^{-\frac{j2\pi n i}{N}}$ and $e^{\frac{j2\pi n i}{N}}$ are the corresponding basis functions of DFT/IDFT. According to Sec.~\ref{sec:Mathematical}, we can achieve such mathematical calculations with the designed neural network models as in Fig.~\ref{fig:math-to-nn}. For a neural network-based DFT model, the kernels of the transposed convolutional layer are set to the real and imaginary parts of $e^{-\frac{j2\pi n i}{N}}$. Similarly, the NN-based IDFT model has the kernels derived based on $e^{\frac{j2\pi n i}{N}}$.


\paragraph{Convert QAM Demapper/Mapper to neural network}
QAM mapper operates like the look-up table, which maps bits to complex symbols based on the determined constellation diagrams in the protocols. The easiest way to implement QAM Demapper/Mapper based on neural networks is to use the \textit{Embedding layer} in nn or \textit{$torch.masked\_select()$} to implement it. Load the predefined constellation parameters in \textit{IEEE 802.11} into the Embedding dictionary or the weight of the linear layer in advance. Implement "lookup table operation". Unfortunately, structures involving index operations are often non-differentiable, and there is no way to pass gradients, which will bring huge disaster to the End-to-end structure.
To achieve differentiable operations, we managed to build a floating-point one-hot vector that supports gradient transfer, and then operate it with a linear layer with standard constellation point weight parameters to achieve differentiable Mapper or DeMapper operations.

\paragraph{Convert quantization process to neural network}
The quantization process is crucial because it restricts the results of the DFT to standard constellation points to ensure that we can derive the corresponding data bits. We use basic operations in PyTorch to implement the corresponding QAM demodulation step by step. The difference is that we embed neural network layers within the standard steps to add learning capabilities.

\begin{figure*}[!tbp]
    \centering
    \includegraphics[width=0.98\linewidth]{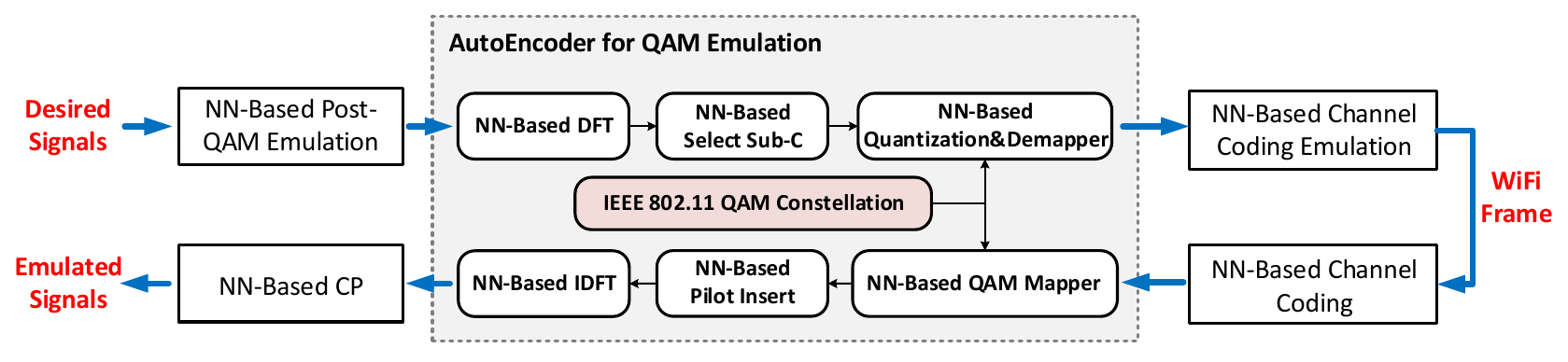}
    \caption{Autoencoder for QAM emulation}
    \label{fig:autoencoder}
    \vspace{-3mm}
\end{figure*}
\subsection{QAM Emulation via Neural Network}
After converting the processes into different neural network models, we can stack these models to form an Autoencoder-like emulation model as illustrated in Fig.~\ref{fig:autoencoder}. Because the goal of the autoencoder is to reconstruct the input at the output, we can employ the model to conduct different kinds of emulation processes by configuring the different inputs. More specifically, if we feed the signal waveforms as the input and train the autoencoder emulation model to reconstruct the signal waveforms at the output, we can achieve the analog emulation. 
Within the autoencoder emulation model, the NN-based DFT/IDFT and the NN-based QAM mapper have deterministic parameters which are configured based on the mathematical models and standards while the parameters inside the NN-based quantizer are trainable, which intends to capture the useful features to select the proper constellation points for better performance than the manual selection. 

Now let's zoom in on the most important step in Fig.~\ \ref{fig:autoencoder}: Quantification. Quantification determines what kind of WiFi signal we generate to simulate the required ZigBee signal. Our NNCTC can easily implement analog emulation in traditional CTC.

\paragraph{NN-Based Analog Emulation.} As shown in Fig.~\ref{fig:autoencoder}, the QAM emulation is designed as an autoencoder structure. The main idea of emulation is to generate a time domain signal that is equal to the desired signal as much as possible. We assume that $u(t)$ is the desired time domain signal and $v(t)$ is the simulated signal generated by the autoencoder used for QAM emulation. The corresponding discrete forms of $u(t)$ and $v(t)$ signals are u[n] and $v[n]$ respectively, where n is the sampling point. Suppose we use the MSELoss function to calculate the loss value, then the average loss value for $u[n]$, $v[n]$ of N sampling points can be expressed as: 
\begin{equation}
    \centering
    {\rm{Loss}}\left( {{\rm{u}}\left[ {\rm{n}} \right],{\rm{v}}\left[ {\rm{n}} \right]} \right) = \frac{1}{N}\sum {\left( {u\left[ n \right] - v\left[ n \right]} \right)^2},
    \label{eq:autoencoder_loss}
\end{equation}
where N is the total number of sampling points. We assume that $U[n]$ and $V[n]$ are the DFT operation results corresponding to $u[n]$ and $v[n]$ respectively. According to Parseval’s theorem, the total energy of the signal in the time domain is equal to the total energy of the signal in the frequency domain. So we have:
\begin{equation}
    \centering
   \mathop \sum \limits_{n = 0}^{N - 1} {\left| {u\left[ n \right] - v\left[ n \right]} \right|^2} = \frac{1}{{\rm{N}}}\mathop \sum \limits_{k = 0}^{N - 1} {\left| {U\left[ k \right] - V\left[ k \right]} \right|^2}.
    \label{eq:Parseval_s_theorem}
\end{equation}
\begin{equation}
    \centering
   {\rm{Loss}}\left( {{\rm{u}}\left[ {\rm{n}} \right],{\rm{v}}\left[ {\rm{n}} \right]} \right) = \frac{1}{{{N^2}}}\mathop \sum \limits_{k = 0}^{N - 1} {\left| {U\left[ k \right] - V\left[ k \right]} \right|^2}.
    \label{eq:autoencoder_loss_to_freq}
\end{equation}
Therefore, combining equation \ref{eq:autoencoder_loss} and equation \ref{eq:Parseval_s_theorem}, we can get the relationship between the loss value and the frequency domain signal, that is, equation \ref{eq:autoencoder_loss_to_freq}. According to the equation \ref{eq:autoencoder_loss_to_freq}, we know that gradient descent is performed based on the loss value in the neural network, which ultimately reduces the absolute value of the difference between the desired frequency domain component $U[n]$ and the simulated frequency domain component $V[n]$. When reflected on the constellation diagram, the nearest constellation point is selected, so we can implement emulation through neural networks.

\paragraph{NN-Based Digital Emulation.}
\begin{figure}[!tbp]
    \centering
    \subfloat[Analog Emulation]{\includegraphics[width=0.45\linewidth]{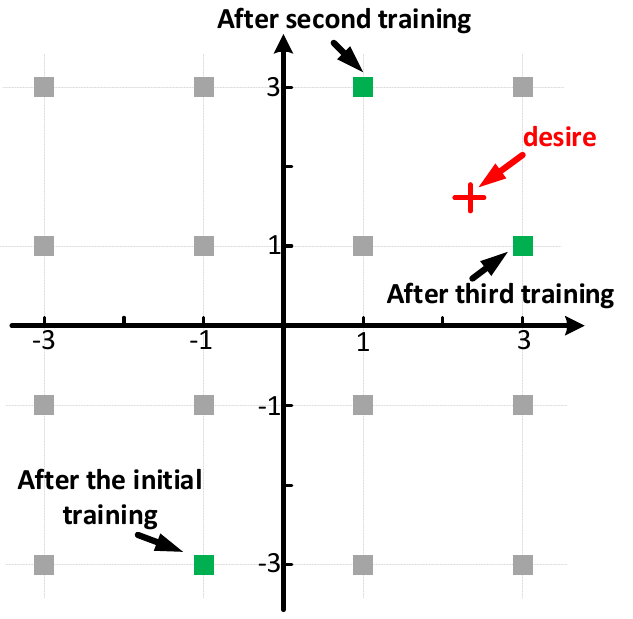}} 
    \subfloat[Digital Emulation]{\includegraphics[width=0.54\linewidth]{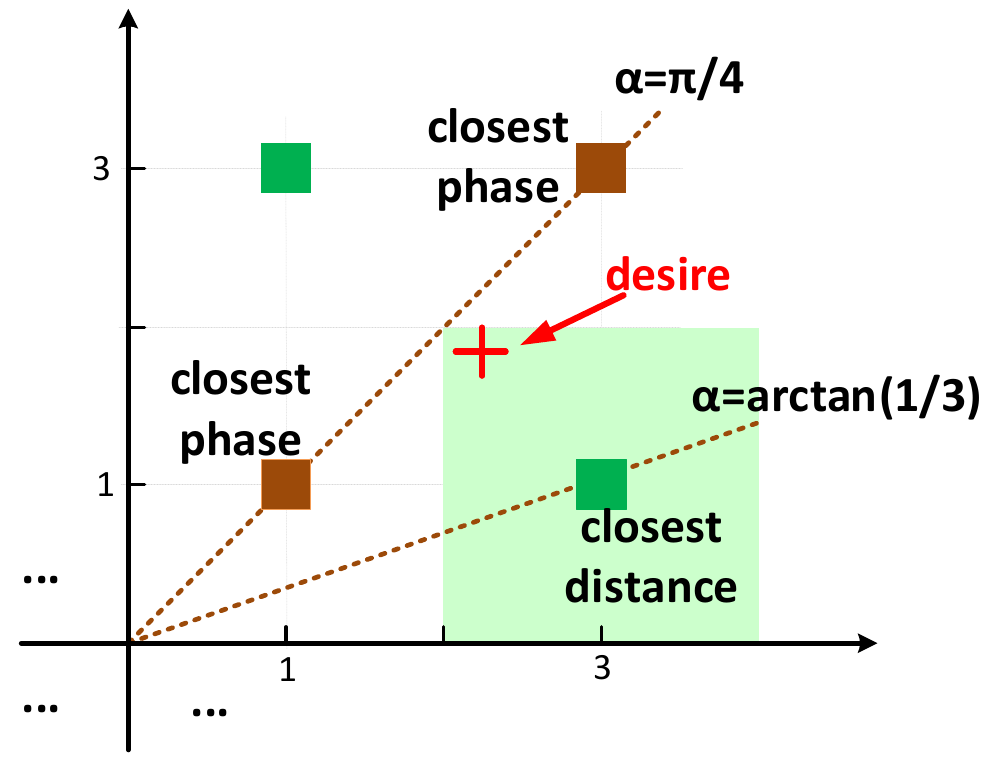}} 
    \caption{Constellation Diagram Comparison of Analog/Digital Emulation}
    \label{fig:Constellation_analog_digital}
    \vspace{-3mm}
\end{figure}
As shown in Fig.~\ref{fig:Constellation_analog_digital}, there are differences in the selection of constellation points between analog emulation and digital emulation. In addition to analog simulation, we can make minor modifications to the implemented NNCTC model to adapt to the idea of digital emulation. ZigBee uses phase difference to demodulate, so some errors may occur in the emulation. As shown in Fig.~\ref{fig:Constellation_analog_digital}~(b), according to the emulation strategy, selecting the closest constellation point will cause the phase error to become larger, which may lead to a reduction in the accuracy of ZigBee demodulation. Specifically, when training the QAM simulation module, we avoid using time-domain signals as input and instead utilize phase signals. This approach allows the model to progressively approximate the target signal in the phase domain under the constraints of the loss function, ultimately realizing the concept of digital emulation.
We still assume that $u[n]$ and $v[n]$ represent the desired time domain signal and the simulated time domain signal respectively. The phase corresponding to $u[n]$ is $h[n]$, and the phase corresponding to $v[n]$ is $q[n]$. Then the mean square error loss value (MSELoss) can be written as:
\begin{equation}
    \centering
   {\rm{Loss}}\left( {{\rm{h}}\left[ {\rm{n}} \right],{\rm{q}}\left[ {\rm{n}} \right]} \right) = \frac{1}{N}\sum {\left( {h\left[ n \right] - q\left[ n \right]} \right)^2},
    \label{eq:autoencoder_loss_digital}
\end{equation}
where N is the number of signal sampling points. Under the guidance of the loss function, the neural network will use the gradient descent method to minimize the difference between $h[n]$ and $q[n]$. Ultimately, the phase of the signal generated by the emulation is close to the phase of the desired signal.

\section{NN-Based POST-QAM Emulation}\label{section:post-qam-channel-coding}\label{section:post-qam-channel-coding}
To incorporate the CTC into the normal Wi-Fi transmission path, we need further processing after the QAM emulation. More specifically, the encoding process before OFDM modulation and some other mechanisms in OFDM transmission have severe effects on emulated signals for CTC. We propose to use neural networks to encounter these effects, which are denoted as NN-based post-QAM emulation and NN-based channel coding emulation.

\subsection{NN-Based Post-QAM Emulation}
\paragraph{Issues in Post-QAM Emulation}
Post-QAM emulation is designed to handle the inherent emulation distortion encountered by OFDM modulation, which is majorly caused by the cyclic prefix introduced in the OFDM modulator. 

\begin{figure}[tbp]
    \centering
    \includegraphics[width=1\linewidth]{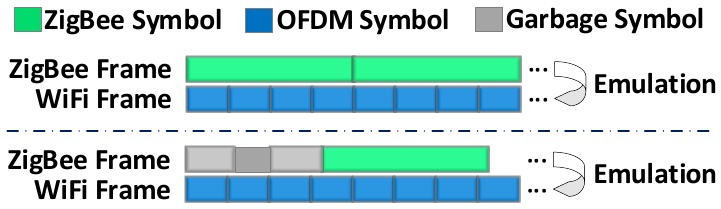} 
    \caption{Alignment of WiFi and ZigBee Symbols}
    \label{fig:post-qam-emulation}
    \vspace{-3mm}
\end{figure}

During the emulation process, \textit{One} ZigBee symbol ($16$us) is segmented into $4$ fragments, each of which has the same $4$us duration as \textit{one} Wi-Fi frame, as illustrated in Fig.~\ref{fig:post-qam-emulation}. 
However, ZigBee symbols do not necessarily need to align with WiFi symbols (or OFDM symbols), as illustrated in the lower part of Fig.~\ref{fig:post-qam-emulation}. In this case, due to the presence of garbage symbols in the ZigBee frame, subsequent ZigBee symbols may not align with OFDM. However, this does not impede the emulation process for NNCTC. This is because NNCTC processes waveforms continuously during the emulation, rather than simulating individual symbols one by one. Therefore, it is not mandatory to achieve symbol alignment. Unfortunately, due to the CP removal procedure, part of the ZigBee segments will not be processed in the QAM emulation, which leads to the inherent post-QAM emulation distortion.

To reduce the impact of cyclic prefixes, WEBee~\cite{li2017webee} relies on the error correction capability of ZigBee to counter the post-QAM emulation distortion, with a selective boundary-flipping strategy. 
And the remaining errors will be corrected thanks to the DSSS scheme in the ZigBee receiver. However, the proposed selective boundary-flipping strategy still suffers severe distortion effects and introduces extra complexity on the transmitter side.

\paragraph{Other Challenges Posed by Post-QAM Emulation}
In addition to the cyclic prefix (CP), Post-QAM emulation also needs to deal with the impact of pilot subcarriers and the selection of data subcarriers. The symbols on pilot subcarriers are fixed for channel tracking, so we cannot manipulate these pilot subcarriers to transmit the CTC load. Therefore, we must carefully select the data subcarriers.  Because of the existence of pilot subcarriers, we must carefully select the subcarriers.

\paragraph{Post-QAM Emulation Implemented by Neural Network}

\begin{figure}[!tbp]
    \centering
    \includegraphics[width=1.0\linewidth]{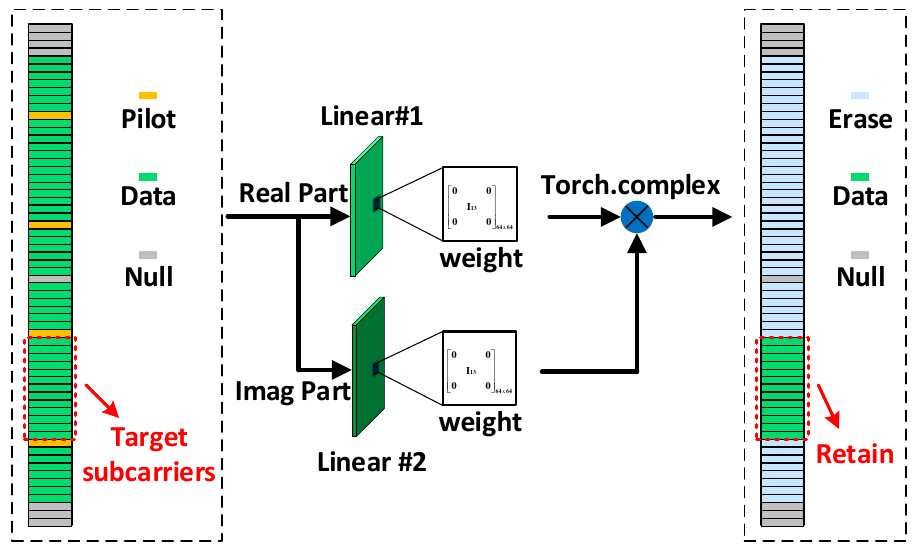}
    \caption{A Case: Select subcarriers using fully connected layers}
    \label{fig:Select-subc-using-fc-layers}
    \vspace{-3mm}
\end{figure}
In NNCTC, we can reduce signal emulation distortion in a manner similar to traditional manual techniques while also supporting the construction of neural networks for cyclic prefix (CP) processing and subcarrier selection to form an end-to-end structure. More specifically, we model the CP add and removal process through linear layers. The corresponding weight of linear layers, $\mathbf{W}_{A}$ and $\mathbf{W}_{R}$, is configured as in Equation~\ref{eq: CP_weight}, where $\mathbf{I}_{[\cdot]}$ represents the identity matrix and $\mathbf{0}$ represents the matrix with all-zero elements. Furthermore, we append other linear layers to selectively select the data subcarriers.

\begin{equation}
    \mathbf{W}_{A} = \left [\begin{array}{c}
        \mathbf{0} \quad \mathbf{I}_{16} \\
         \mathbf{I}_{64}
    \end{array} \right ]_{80\times 64}, \quad \mathbf{W}_{R} = \left [\begin{array}{c}
        \mathbf{0} \quad \mathbf{I}_{64}
    \end{array} \right ]_{64\times 80}
    \label{eq: CP_weight}
\end{equation}

It's worth pointing out that the linear layers for CP processing and subcarrier selection are configured with the fixed weight, and they won't be trained. For example, Fig.~\ref{fig:Select-subc-using-fc-layers} illustrates a basic scenario, demonstrating how the subcarrier selection process is accomplished in a neural network. Assuming the desired ZigBee components are already located in the "Target subcarriers" in the figure, we then use two linear layers to separately process the real and imaginary parts of the signal. The strategy outlined by Equation~\ref{eq: CP_weight} guides the appropriate configuration of the weights in the linear layers, thereby retaining only the necessary subcarriers. After the linear layers process the signal, the real and imaginary parts are recombined using the $Torch.complex$ operation, effectively eliminating any extra frequency components. The main purpose of NN-based post-QAM emulation is to integrate these processes into the end-to-end pipeline for CTC emulation. Therefore, the autoencoder can be trained to reduce the distortion introduced by CP processing and find the proper QAM constellations that match the requirements.

To enable the CTC transmission by crafting the binary payload, we are supposed to derive the bits based on the derived symbols. As depicted in Fig.~\ref{fig:NNCTC_Overview}, within the WiFi transmission path, the binary payloads are encoded by a channel encoder, and the encoded bits are mapped to the QAM symbols. Therefore, once we infer the symbols from the QAM emulation model, we can derive the binary payloads based on the QAM mapping and the channel coding schemes. To achieve this, we follow the recognized idea in WEBee by solving the linear function on GF(2) as in Equation~\ref{eq:gf(2)}, where $X$ and $Y$ represent the binary payloads and encoded bits and $G$ is the coding matrix for channel coding schemes. 

\begin{equation}
\centering
    \mathrm{Y}=\mathrm{G}\times_{\mathrm{GF(2)}}\mathrm{X}.
\label{eq:gf(2)}
\end{equation}

\section{Evaluation}\label{sec: evaluation}
In this section, we will conduct a substantial number of experiments to evaluate NNCTC. Furthermore, we will showcase the potential of NNCTC to optimize traditional CTC. NNCTC can support commercial WiFi devices, allowing the generated WiFi PSDU to be modulated with the standard WiFi physical layer to generate waveforms.
\begin{figure}[!tbp]
    \centering
    \includegraphics[width=0.90\linewidth]{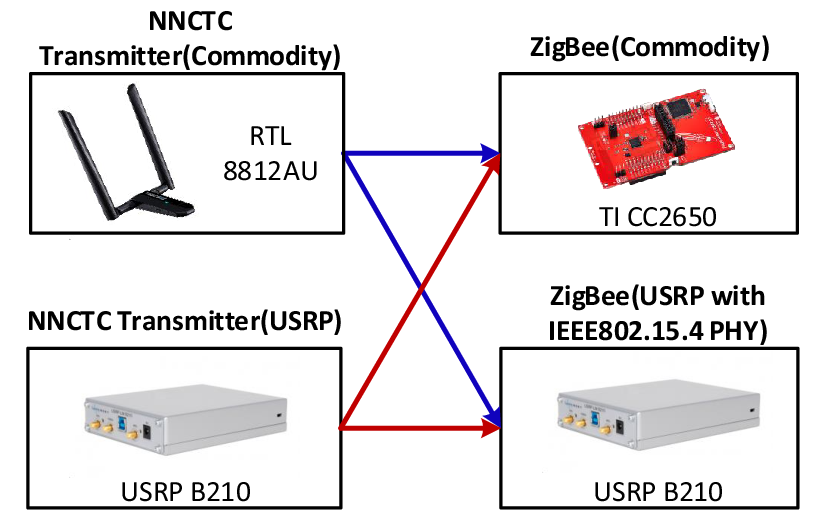}
    \caption{The Experimental Transmitter Devices}
    \label{fig:the-transmitter-devices}
    \vspace{-3mm}
\end{figure}

\subsection{Experiment Setup}
\paragraph{Hardware and Software Description}
In terms of neural networks, we implemented the designed NNCTC based on PyTorch and ran it on a laptop with x86 architecture. At the same time, an NVIDIA RTX3060 graphics card with 12G memory was used to train the designed NNCTC. At the level of CTC evaluation, we use USRP B210 with IEEE802.11 b/g PHY and commercial WiFi network card RTL 8812AU as the sending end of NNCTC(As shown in the Fig.~\ref{fig:the-transmitter-devices}). USRP B210 with IEEE802.15.4 PHY and TI CC2650 are used as receivers, whereas USRP B210 is only used for evaluation purposes, which can help us obtain low-level physical layer information (such as symbol error rate, etc.).
\begin{figure}[!tbp]
    \centering
    \subfloat[Indoor Scene]{\includegraphics[width=0.50\linewidth]{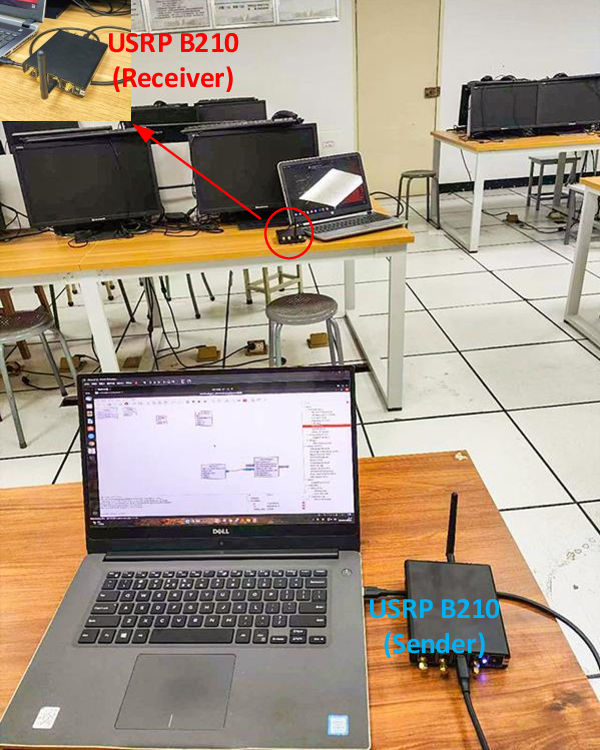}}
    \subfloat[Outdoor Scene]{\includegraphics[width=0.497\linewidth]{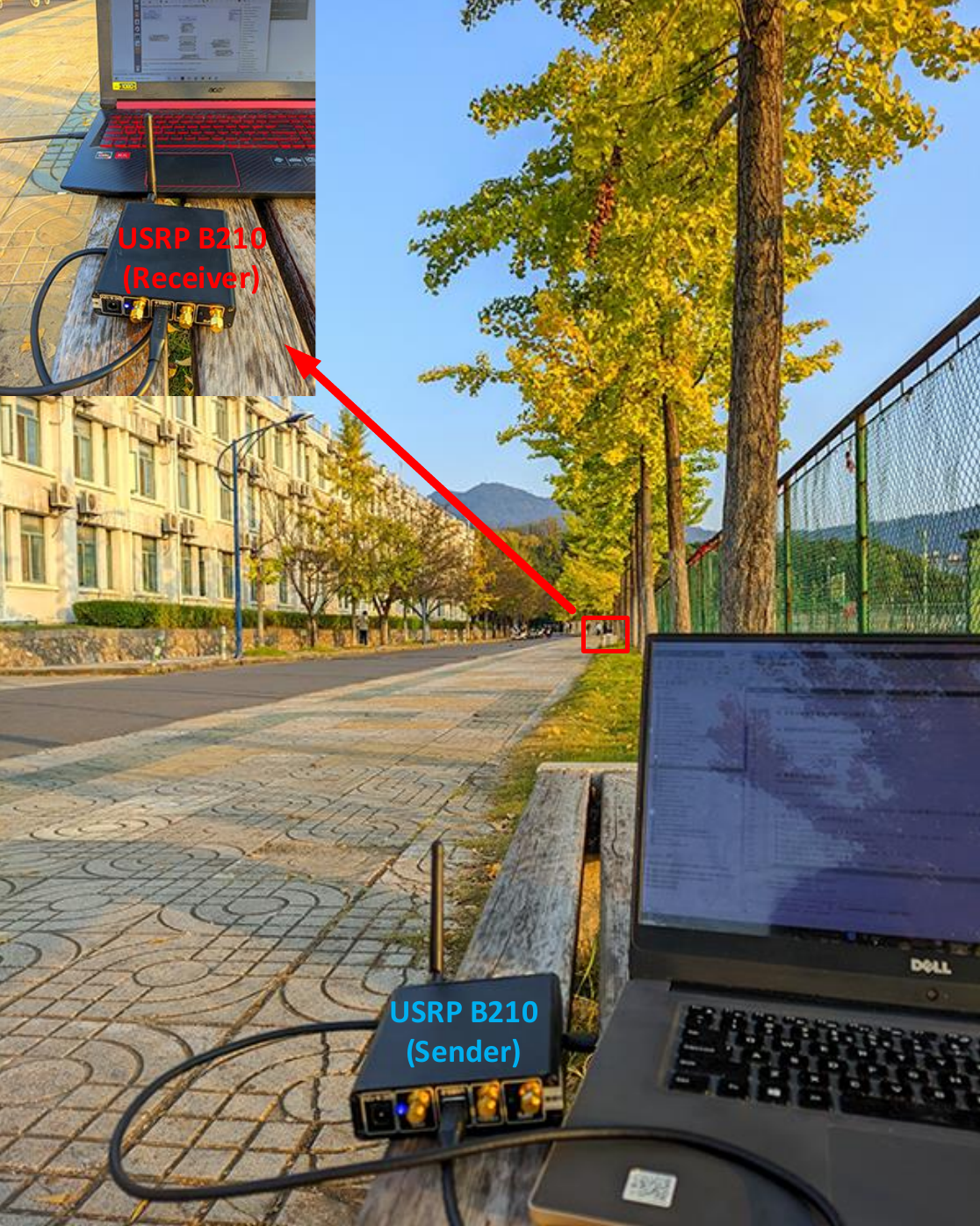}} 
    \caption{Experimental Scene Setup}
    \label{fig:Experimental-scene-setup}
    \vspace{-3mm}
\end{figure}

\paragraph{Experimental Frame Design}
We first conduct an overall performance evaluation of NNCTC and then repeat each experiment ten times to calculate the average values. Specifically, the experiments cover different transmission distances, payload lengths, and transmit powers in indoor and outdoor scenarios. Subsequently, we demonstrate the effectiveness of NNCTC in optimizing traditional CTC. To demonstrate the feasibility of NNCTC, we separately evaluate the core submodules in NNCTC.

\subsection{Overall Performance Evaluation}
In this section, we will conduct a comprehensive evaluation of NNCTC. First, our experimental scene selection is carried out in the indoor scene shown in Fig.~\ref{fig:Experimental-scene-setup}. Secondly, we make the ZigBee payload have 16 symbols length, which includes all ZigBee symbols. The transmit power is set to 10 dBm, and the distance between the WiFi transmitter and the ZigBee receiver is fixed at 4m. Then, under the same experimental settings and interference environment, we compared NNCTC with WEBee and WIDE and evaluated their symbol error types (SER), packet reception rate (PRR), and effective throughput respectively.

The choice of WEBee and WIDE as a basis for comparison is primarily motivated by two factors. Firstly, they both emulate ZigBee signals based on OFDM. Secondly, WEBee and WIDE, respectively representing analog emulation and digital emulation, are classical works in their own right, and these two works are highly representative.

\begin{figure}[tbp!]
    \centering
    \includegraphics[width=0.85\linewidth]{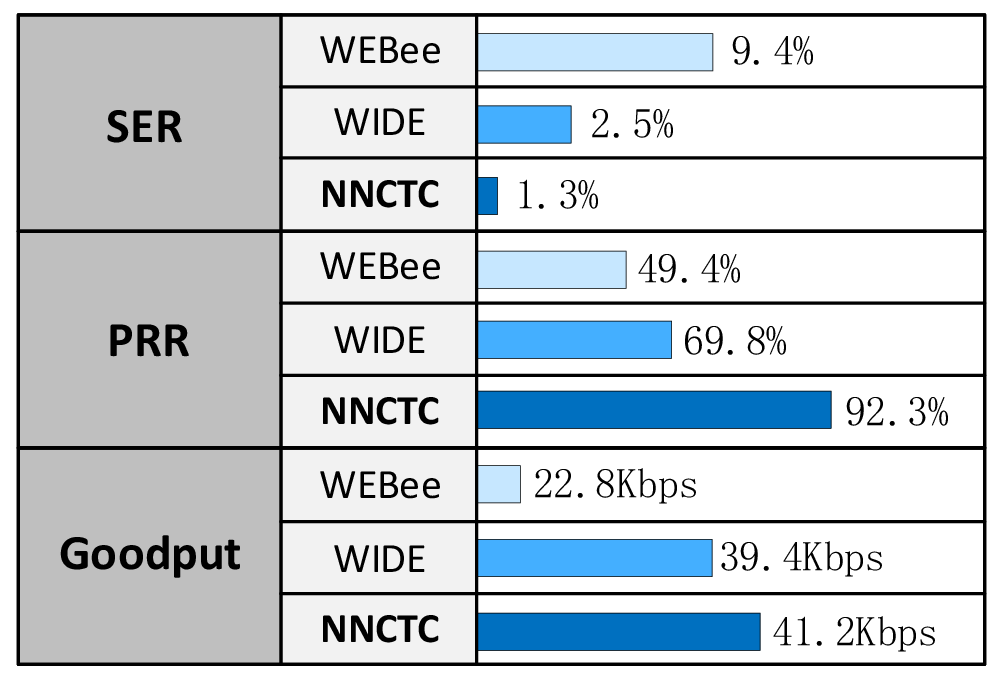}
    \caption{Overall performance comparison}
    \label{fig:overall-performance}
\end{figure}

The comparison results are shown in Fig.~\ref{fig:overall-performance}. The first is the symbol error rate. In the experimental environment introduced in this article, WEBee has the highest symbol error rate, reaching 9.4\%. In comparison, WIDE based on phase emulation and NNCTC proposed in this article have better symbol error rates. 
Since NNCTC in this paper can learn better parameters through manual configuration or training, it can theoretically achieve the emulation strategy's performance limits. 
Secondly, regarding the PRR indicator, see also Fig.~\ref{fig:overall-performance}. Thanks to its flexible parameter configuration, NNCTC can achieve a packet reception rate of up to 96.3\% in the current experimental environment. In comparison, the PRR of WEBee and WIDE are only 49.4\% and 69.8\% respectively. A lower SER can greatly increase the probability of passing preamble detection and data packet synchronization.

Finally, regarding effective throughput, when evaluating the goodput indicator, we only focus on the payload part of the data packet. Note that the overall goodput in our experimental results is smaller than ZigBee's highest rate of 250kbps, which is related to our evaluation experiment settings. During emulation, we choose WiFi frames with fixed PSDU length as carriers to carry CTC signals, so ZigBee signals are not continuously transmitted. However, under the same experimental settings, the Goodput indicator is still valid. Because of NNCTC's higher packet acceptance rate and lower error rate, NNCTC's Goodput indicator (41.2Kbps) is better than the traditional static emulation WEBee and WIDE.

\begin{figure}[!tbp]
    \centering
    \includegraphics[width=0.49\linewidth]{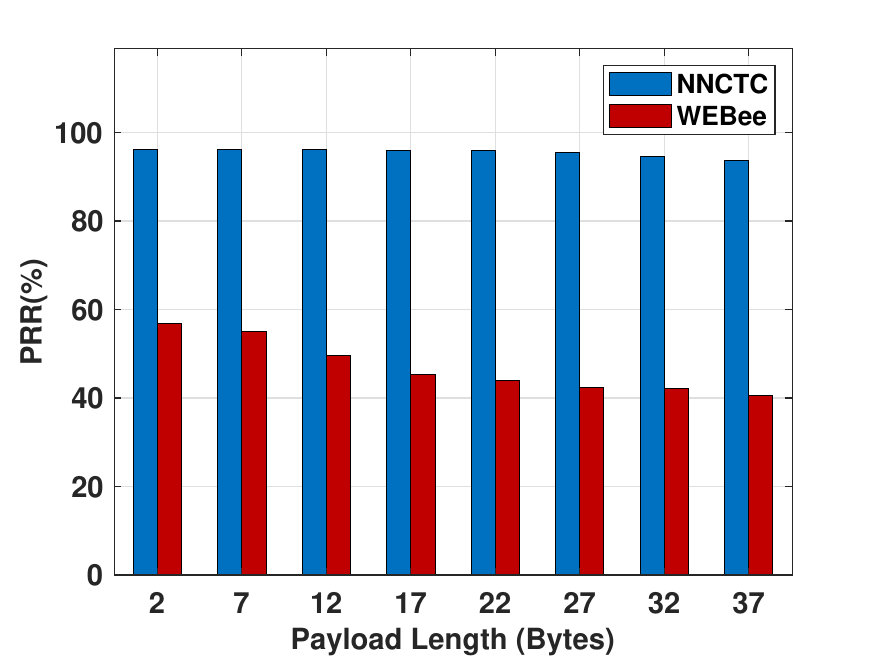}
    \includegraphics[width=0.49\linewidth]{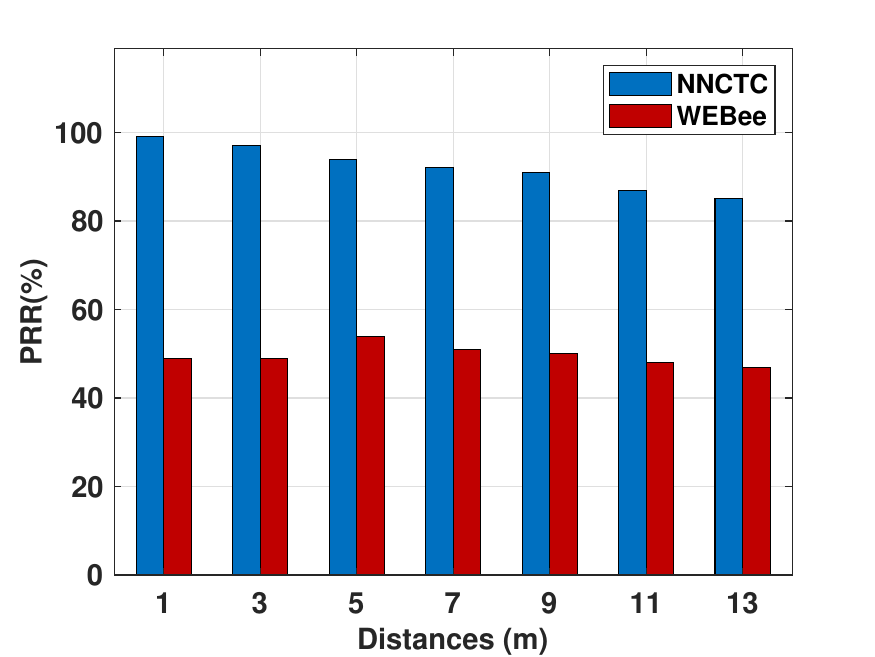}
    \caption{Comparisons under Various Payload Lengths and Transmission Distances}
    \label{fig:prr-payload-distances}
\end{figure}

\paragraph{The effect of different payload lengths}
As shown in Fig.~\ref{fig:prr-payload-distances}, we compared PRR under different payload lengths and transmission distances. We conducted these experiments in an indoor environment with a transmitter transmitting at 10dBm and a receiver located 2.5m away. According to the experimental results, the PRR of WEBee gradually decreases as the payload length increases, whereas our NNCTC exhibits no significant variation. This indicates that NNCTC performs better for longer payloads.

\begin{figure}[!tbp]
    \centering
    \includegraphics[width=1\linewidth]{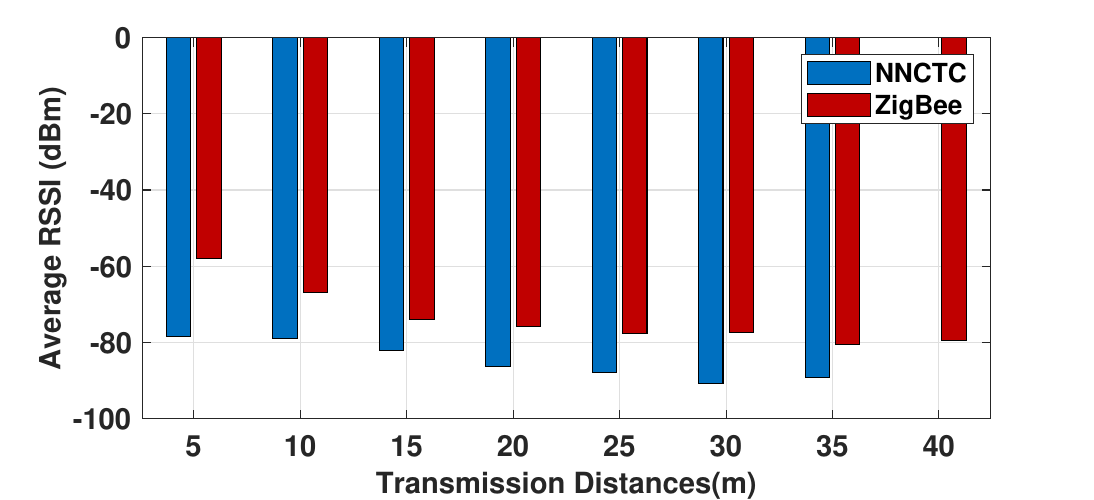}
    \caption{The RSSI of NNCTC and Standard ZigBee at Various Transmission Distances}
    \label{fig:rssi-distances}
    \vspace{-3mm}
\end{figure}

\paragraph{The effects of different distances}
Next, we evaluated the impact of different transmission distances on PRR, keeping the payload length fixed at 32 bytes, and maintaining a constant transmit power of 10dBm. Similarly, refer to Fig.~\ref{fig:prr-payload-distances}. Experimental results indicate that the PRR of WEBee remains below 55\% at various distances we tested. In contrast, our NNCTC exhibits an average PRR exceeding 90\% when the distance is less than 9m. The higher PRR suggests that the emulated ZigBee signals have a lower symbol error rate, making them more likely to pass preamble detection and sync symbol detection. 
In addition to the above, as shown in Fig.~\ref{fig:rssi-distances}, we also compared NNCTC with standard ZigBee in terms of transmission distance. Under the same payload length and transmission power, there is still a certain gap in transmission distance compared to standard ZigBee. However, CTC still has its application scenarios. In CTC, performance degradation is due to incompatible DSP operations in modulation/coding which compromises the native error correction capability in low SNR.

\begin{figure}[tbp!]
    \centering
    \includegraphics[width=0.49\linewidth]{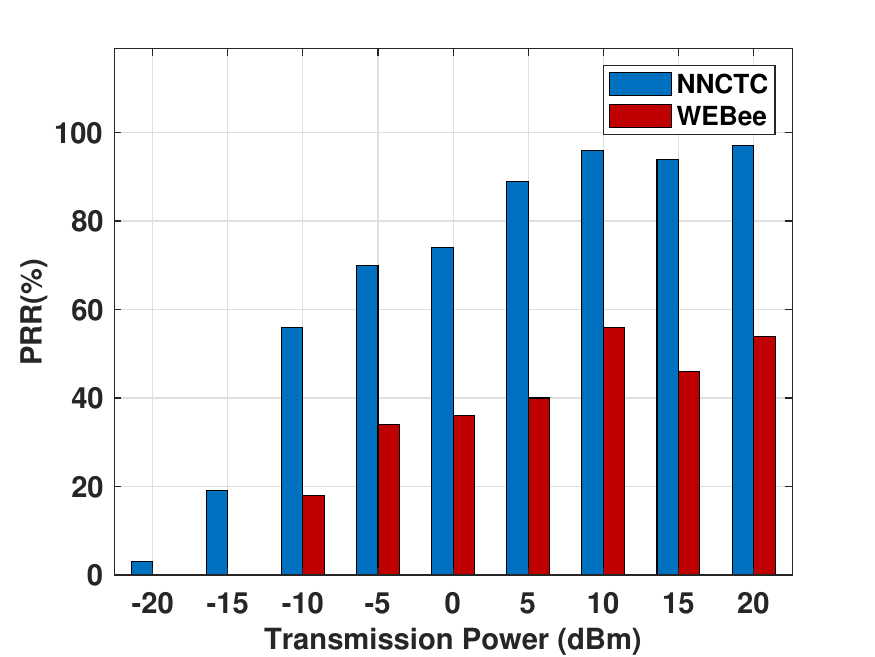}
    \includegraphics[width=0.49\linewidth]{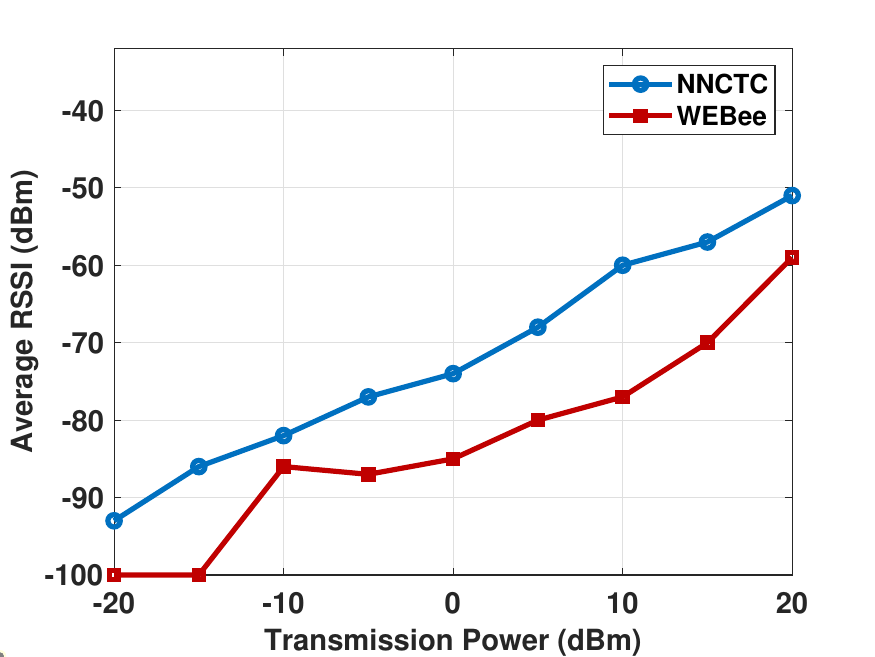}
    \caption{Comparisons under Various Payload Lengths and Transmission Power}
    \label{fig:prr-power-rssi}
    \vspace{-3mm}
\end{figure}

\paragraph{The impact of different transmit powers}
Insufficient transmit power can lead to a higher SNR, resulting in the receiver's inability to demodulate the required signal properly. We evaluated the average PRR at different transmission distances, setting the payload length to 32 bytes, and keeping the transmitter and receiver distance fixed at 2.5m. The experimental results are shown in Fig.~\ref{fig:prr-power-rssi}.
First, from the left figure, we can observe that both WEBee and NNCTC are sensitive to the signal transmit power. When the signal transmit power is as low as -10dBm (equivalent to 0.1mW), WEBee's average PRR is already reduced to around 18\%, while our NNCTC still performs at 56\% PRR. When the transmit power drops to -15dBm, WEBee can no longer function properly, while NNCTC continues to operate.
Furthermore, the right figure illustrates the average RSSI performance at the receiver under different transmit powers, using -100dBm to indicate that ZigBee cannot receive data packets.

\subsection{The traditional CTC optimized by NNCTC}
\begin{figure}[!tbp]
    \centering
    \includegraphics[width=1\linewidth]{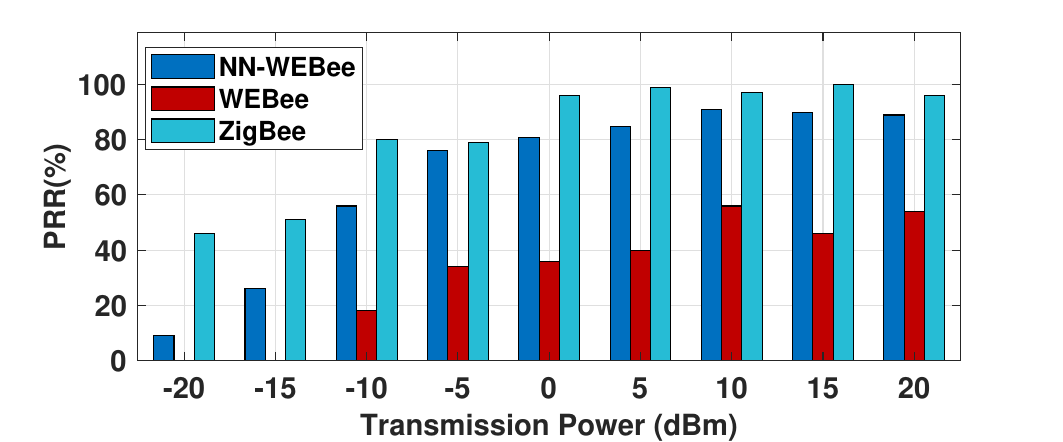}
    \caption{WEBee improved based on NNCTC}
    \label{fig:nn-webee-prr}
    \vspace{-3mm}
\end{figure}
Because NNCTC follows a white-box design philosophy, we can extract certain network model parameters to optimize traditional CTC. With the assistance of NNCTC, we found that in traditional emulation strategies, the "quantization" step in QAM emulation plays a crucial role in the quality of generated signals and the PRR at the receiver. One significant task in the "quantization" step is dynamically scaling the frequency domain components of the signal to match standard QAM constellations. With the optimization provided by NNCTC, we modified the scaling factors used in traditional emulations and continued the evaluation. Taking WEBee as an example, we referred to the optimized WEBee as "NN-WEBee". We conducted extensive experiments on CC2650, with a fixed payload length of 32 bytes, and a fixed distance of 2.5m between the transmitter (USRP B210) and receiver. The experimental results are depicted in Fig.~\ref{fig:nn-webee-prr}. At various transmit power levels, the PRR performance of NN-WEBee significantly outperforms that of traditional WEBee. These experiments demonstrate that NNCTC has the potential to assist in optimizing traditional CTC.

\subsection{NN-Based QAM Emulation Evaluation}

\begin{figure}[tbp]
    \centering
    \subfloat[NN-Based DFT]{\includegraphics[width=0.45\linewidth]{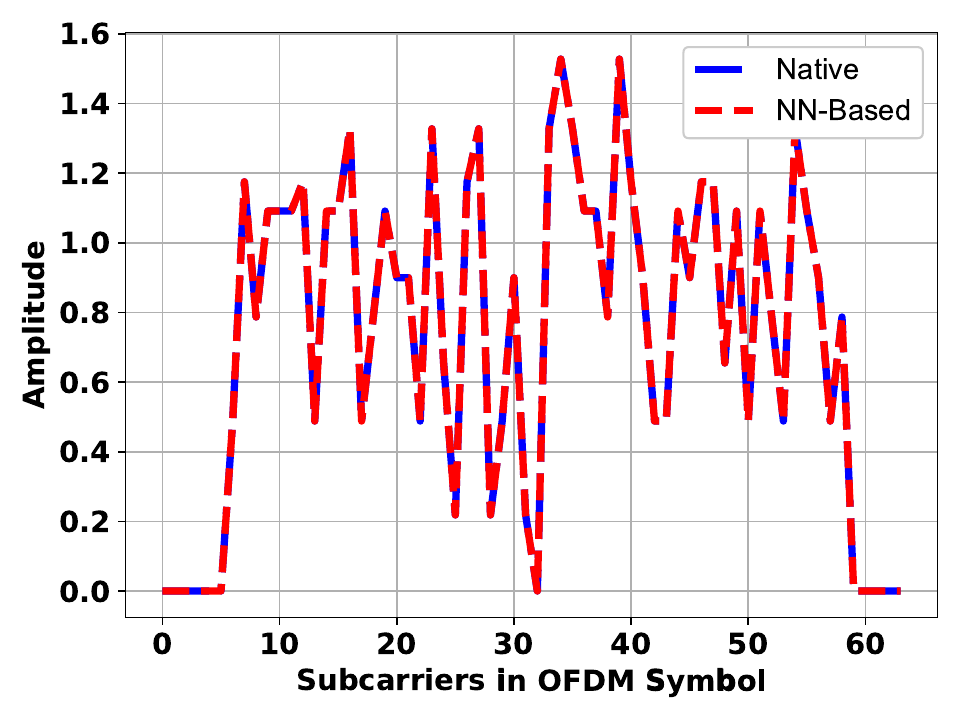}}
    \subfloat[NN-Based IDFT]{\includegraphics[width=0.45\linewidth]{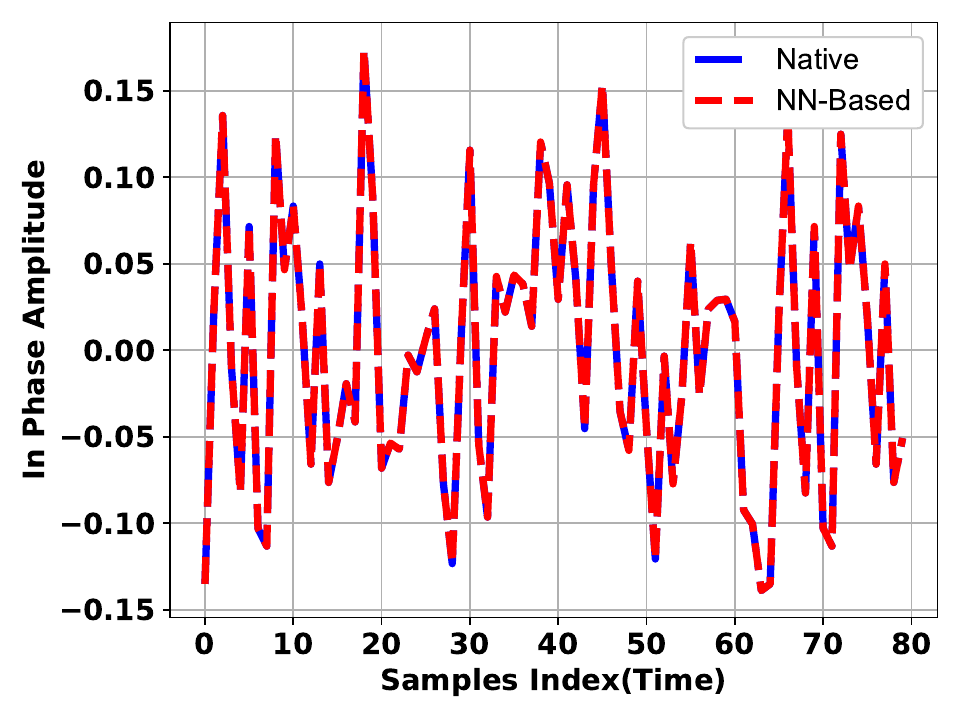}} \\ 
    \subfloat[NN-Based QAM Mapper]{\includegraphics[width=0.45\linewidth]{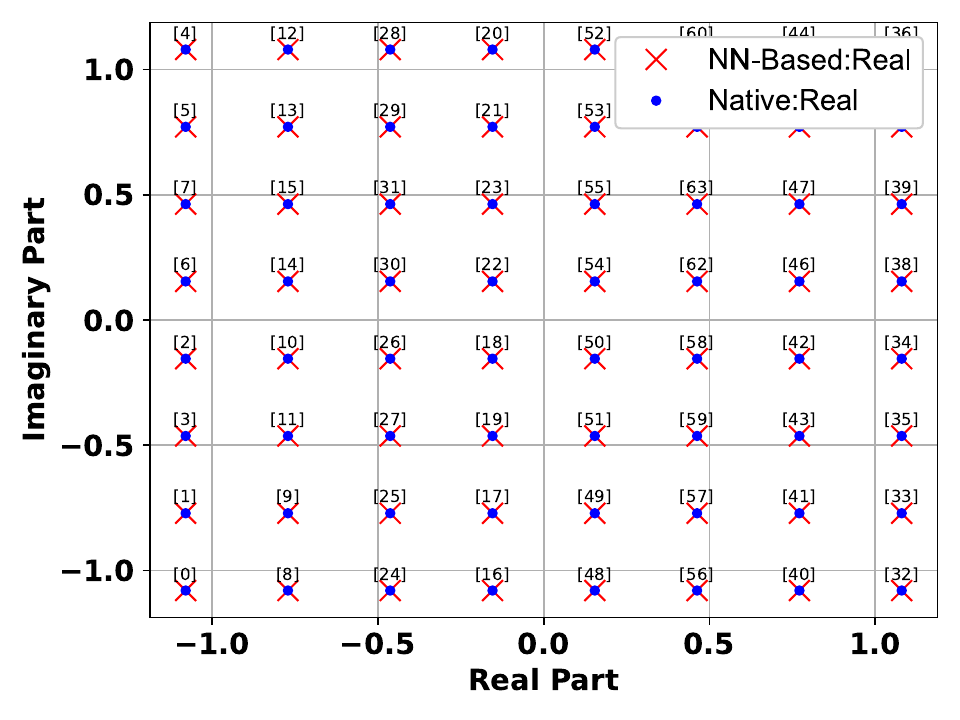}}
    \subfloat[NN-Based QAM Demapper]{\includegraphics[width=0.45\linewidth]{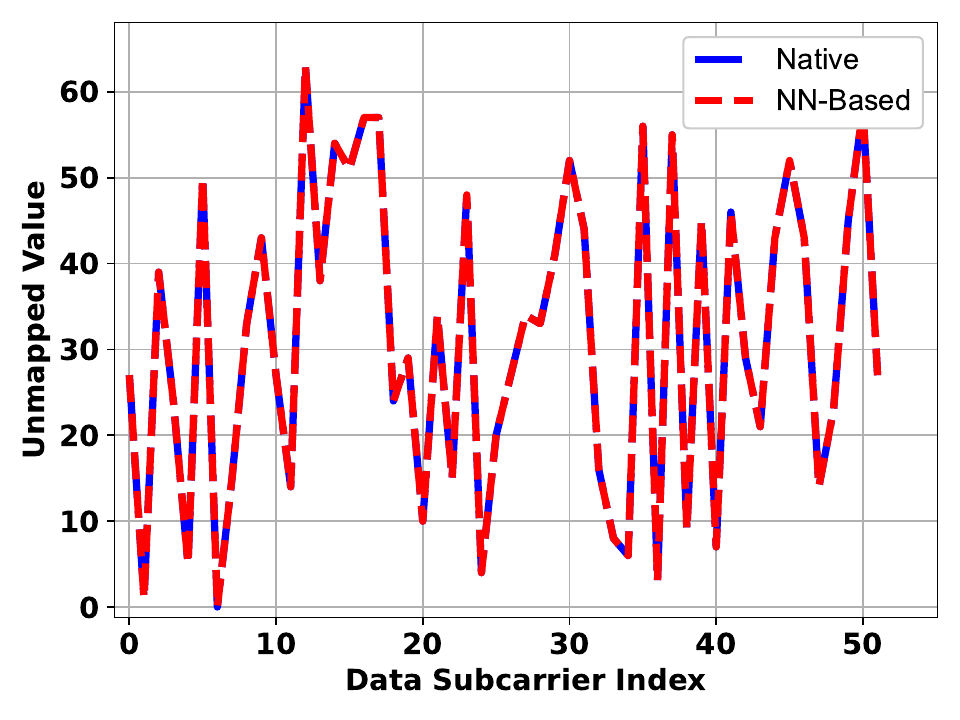}}
    
    \caption{Evaluation of the Core Components Within the Autoencoder}
    \label{fig:autoencoder-inner}
    \vspace{-3mm}
\end{figure}

\begin{figure}[tbp]
    \centering
\includegraphics[width=0.98\linewidth]{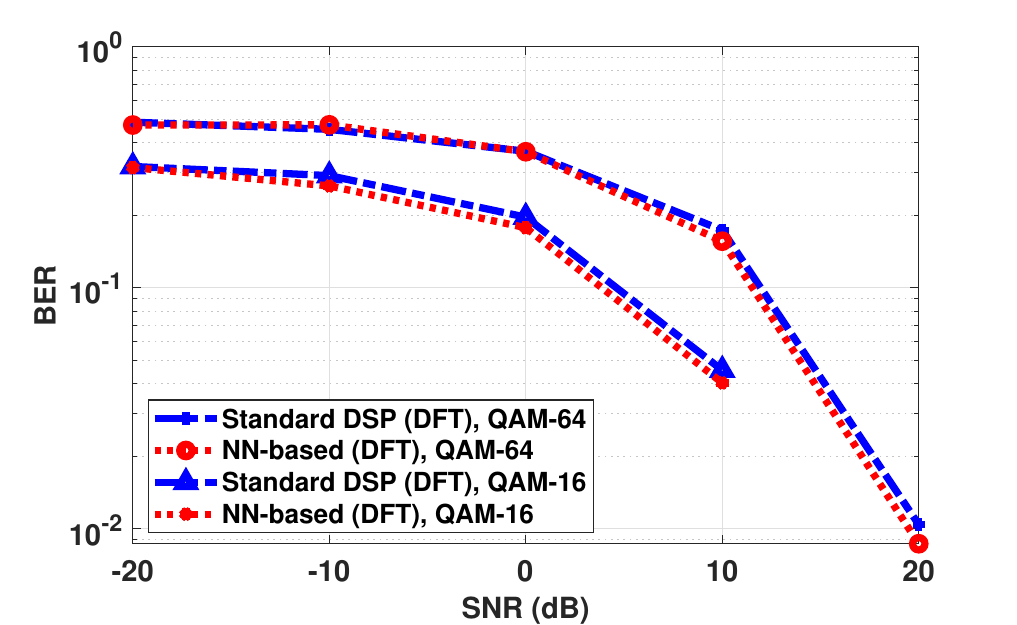}
    \caption{NN-Based DSP vs Standard DSP}
    \label{fig:nnvsdsp_snr}
    \vspace{-3mm}
\end{figure}

    

Based on the standard data (from Matlab), we evaluated several main modules inside the autoencoder, and the results are shown in Fig.~\ref{fig:autoencoder-inner}. The autoencoder for QAM emulation includes four main components, namely DFT, IDFT, QAM Mapper, and QAM Demapper (Quantization). In Fig.~\ref{fig:nnvsdsp_snr}, we evaluate the BER of NN-based components and standard DSP components under various channel conditions. We utilize signals modulated with QAM-16 and QAM-64 for evaluation. Experimental results demonstrate that under different SNR channels, NN-based components perform nearly identically to standard DSP components. Experimental results show that the design based on a neural network is feasible and can achieve the same effect as the standard process.

\begin{figure}[tbp]
    \centering
    \subfloat[In-Phase]{\includegraphics[width=0.48\linewidth]{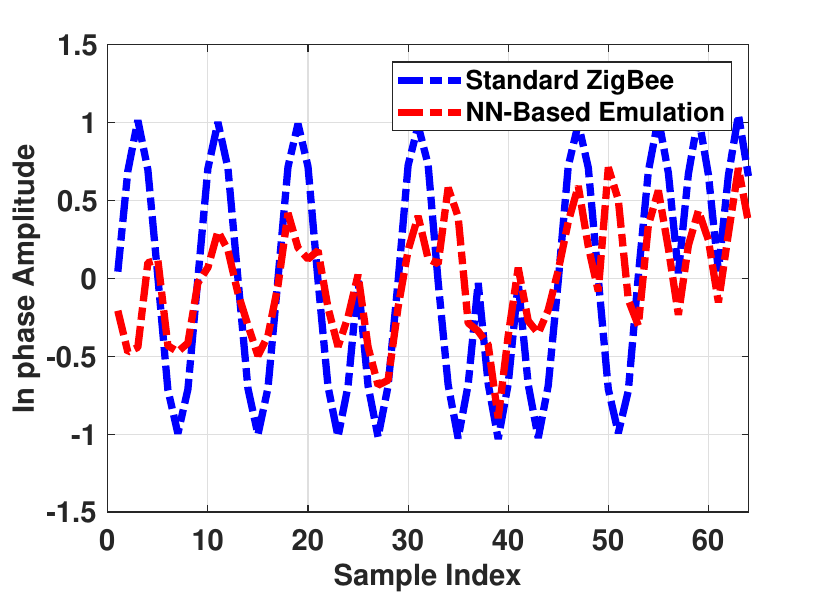}}
    \subfloat[Quadrature]{\includegraphics[width=0.48\linewidth]{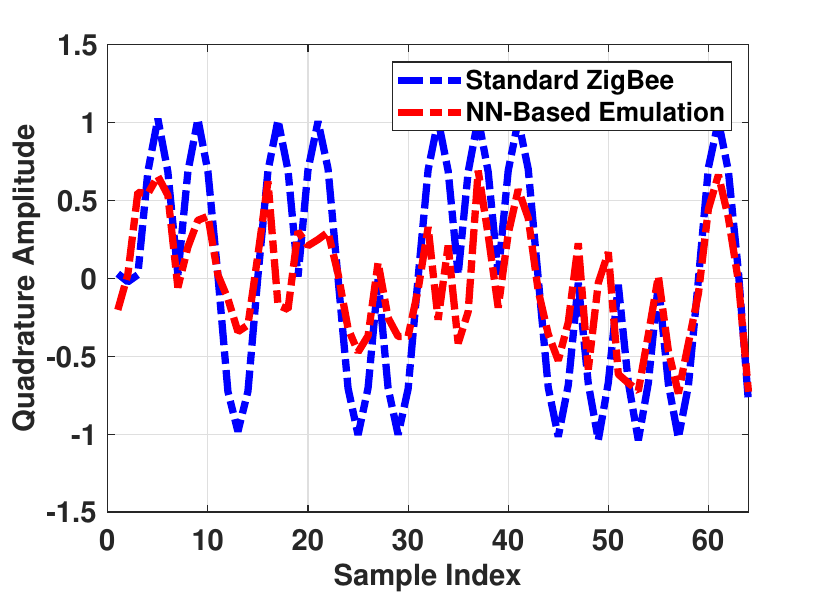}} 
    
    \caption{NN-based QAM Emulation Signal and Original ZigBee Signal}
    \label{fig:qamemulated-nn-result}
    \vspace{-3mm}
\end{figure}

Next, Fig.~\ref{fig:qamemulated-nn-result} shows a schematic diagram of NN-based QAM emulation. The figure shows the difference between the emulation results of 2 ZigBee signals and the original ZigBee signal. Except for the error caused by the cyclic prefix, the remaining part of the signal is almost consistent with the expected signal, which shows the feasibility of NN-based QAM emulation.

\begin{figure}[!tbp]
    \centering
    \subfloat[Before Quantization]{\includegraphics[width=0.48\linewidth]{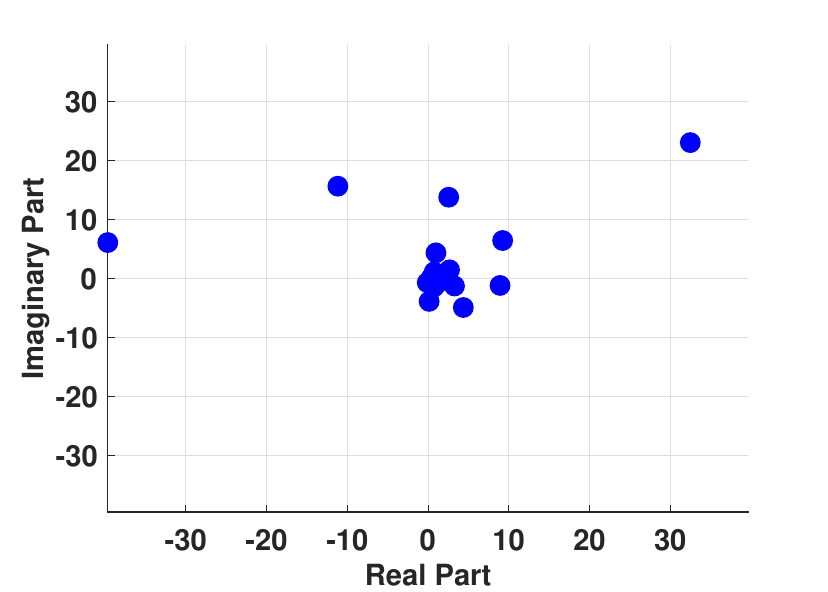}}
    \subfloat[Scaling/Normalization]{\includegraphics[width=0.48\linewidth]{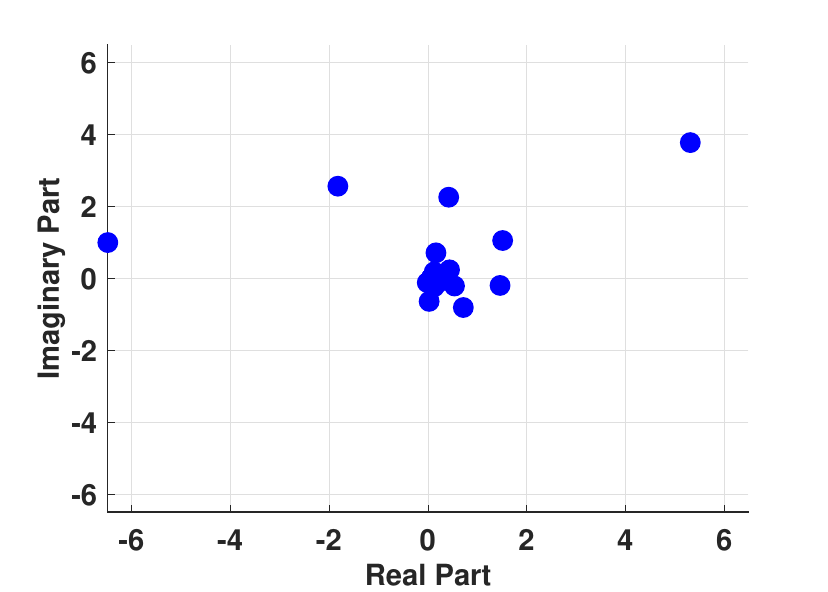}} \\

    \subfloat[Quantization: NNCTC]{\includegraphics[width=0.48\linewidth]{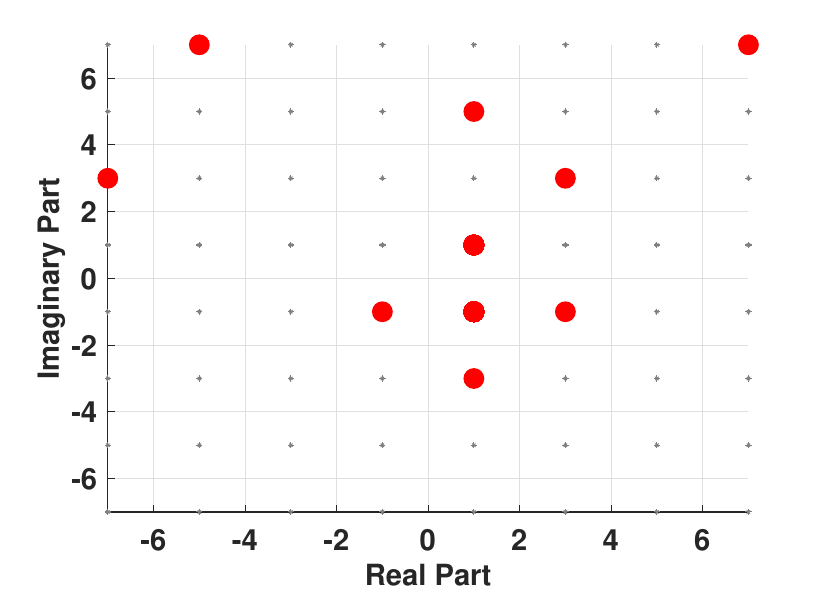}} 
    \subfloat[Quantization: WEBee]{\includegraphics[width=0.48\linewidth]{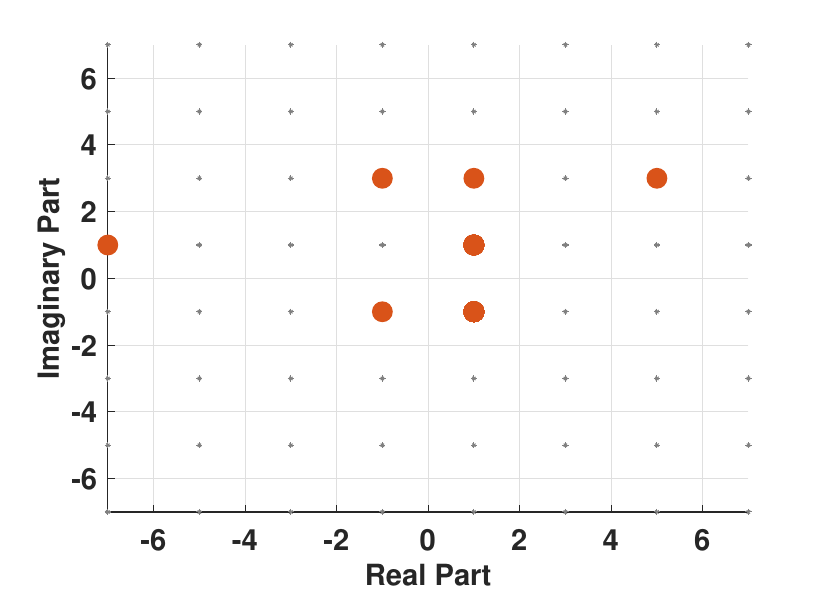}} 
    \caption{Comparison of NNCTC and Traditional CTC Emulations in the Quantization Step}
    \label{fig:Quantization-Step}
    \vspace{-3mm}
\end{figure}

\paragraph{Evaluation of the Quantization module}
Taking a 1/4-rate signal of a ZigBee symbol as an example, after removing the cyclic prefix (CP) and applying DFT, its distribution on the constellation diagram is shown in Fig.~\ref{fig:Quantization-Step}~(a). At this point, we can observe that the frequency domain components of the signal are scattered and the difference between the maximum and minimum values is significant. 

Then, in the case of WEBee, the Quantization process requires manual step-by-step execution. For example, the first step is to normalize the entire frequency domain components to scale them appropriately for the constellation diagram. After normalization, it appears as shown in Fig.~\ref{fig:Quantization-Step}~(b). 
After quantization in WEBee, the distribution of the signal on the constellation diagram appears as shown in Fig.~\ref{fig:Quantization-Step}~(d).

With NNCTC in place, there is no need for manual scaling and constellation point selection. This entire process is delegated to the neural network. When we feed the frequency domain components of the raw signal to the trained NNCTC, after the Quantization process, we directly obtain the standard constellation points. The distribution on the constellation diagram appears as shown in Fig.~\ref{fig:Quantization-Step}~(c). By comparing the (c) and (d) plots in Fig.~\ref{fig:Quantization-Step}~(c), we can observe that the constellation points chosen by NNCTC are more dispersed and have larger amplitudes. This has the advantage of allowing the generated CTC signal to contain more phase information. In contrast, because WEBee manually executed normalization, the distribution of constellation points in the frequency domain signal is concentrated in the middle position, making it susceptible to signal information loss.

\begin{figure}[!tbp]
    \centering
    \subfloat[Original]{\includegraphics[width=0.32\linewidth]{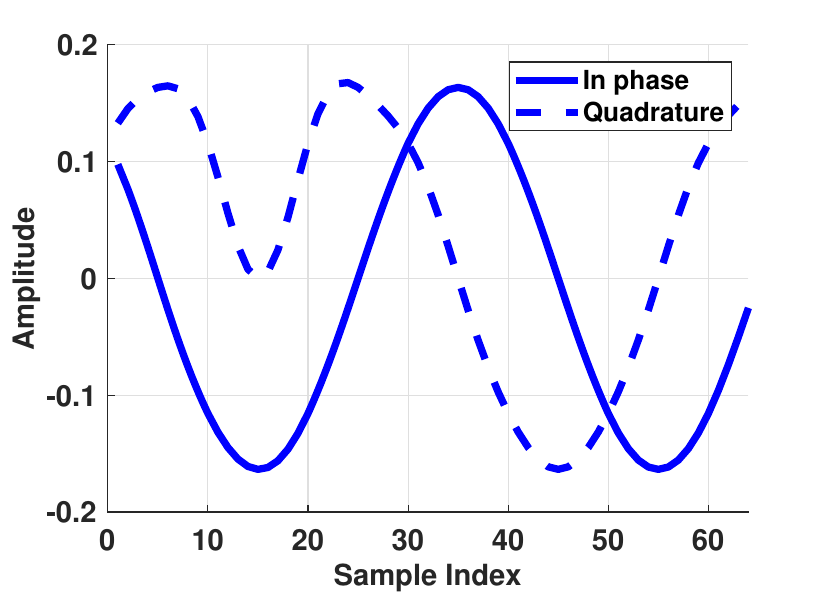}}
    \subfloat[WEBee]{\includegraphics[width=0.32\linewidth]{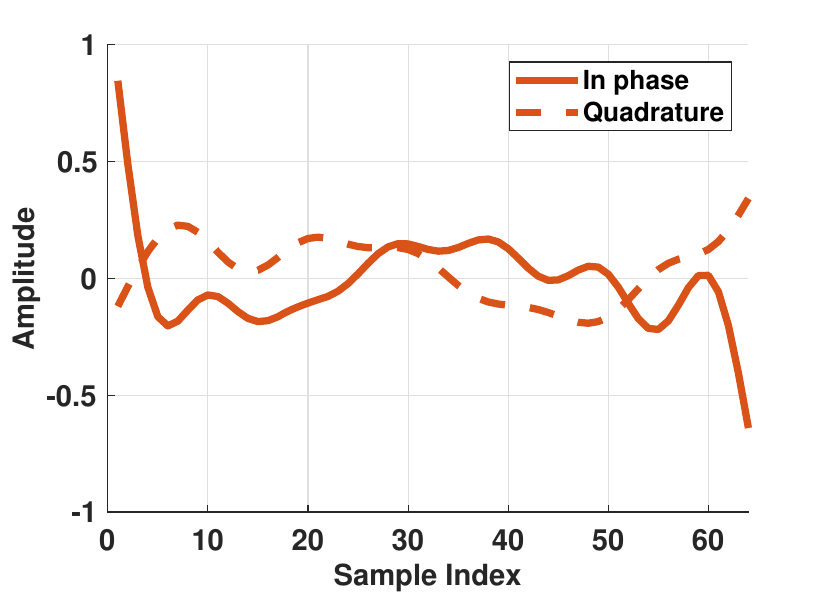}} 
    \subfloat[NNCTC]{\includegraphics[width=0.32\linewidth]{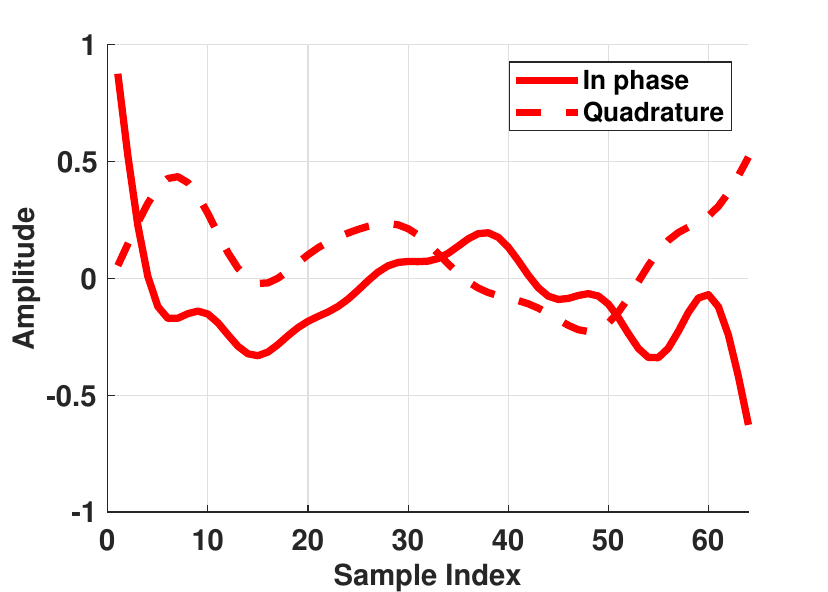}} 
    \\
    \subfloat[Original]{\includegraphics[width=0.32\linewidth]{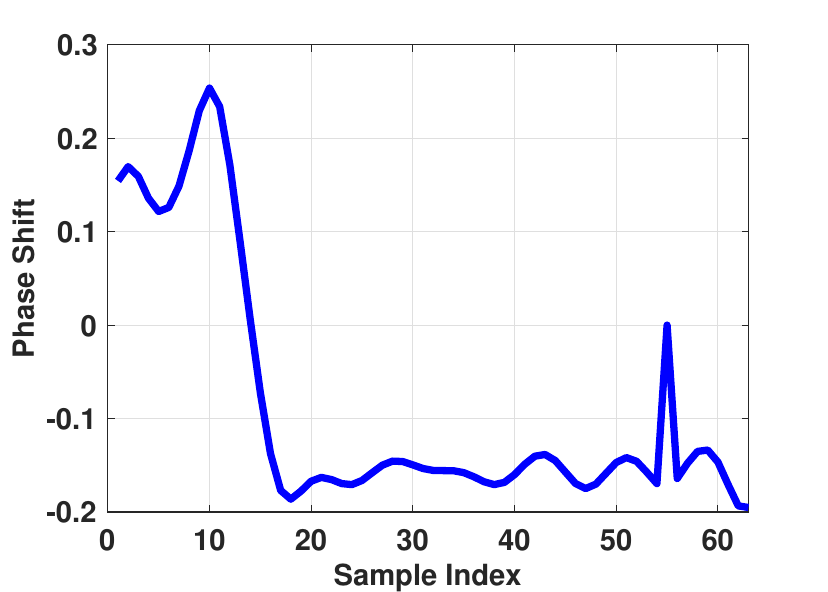}}
    \subfloat[WEBee]{\includegraphics[width=0.32\linewidth]{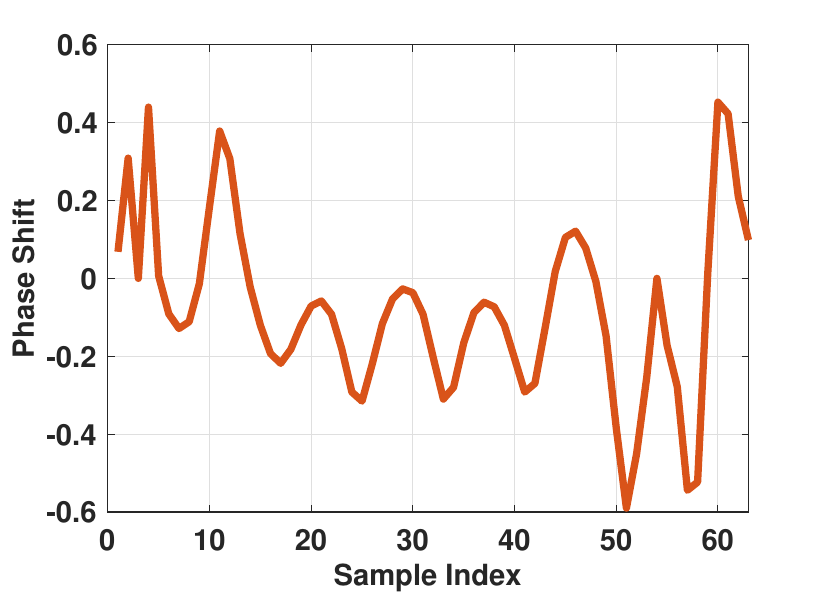}} 
    \subfloat[NNCTC]{\includegraphics[width=0.32\linewidth]{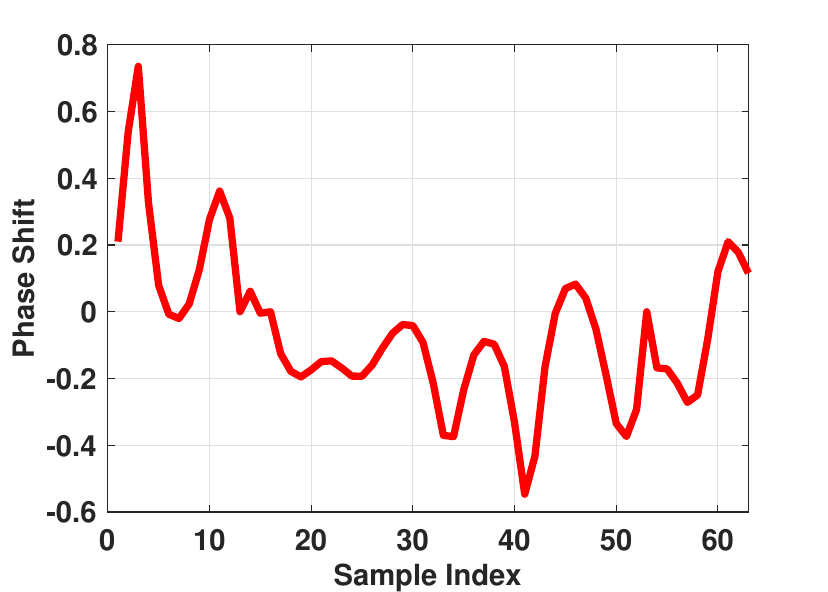}} 

    \caption{After Quantization: Comparison of Phase Differences in Signals}
    \label{fig:Quantization-phase-shift}
    \vspace{-3mm}
\end{figure}

Figure \ref{fig:Quantization-phase-shift} illustrates the time-domain signals and phase shifts generated by NNCTC and WEBee after the Quantization process. Note that this does not include adding the cyclic prefix (CP) and selecting subcarriers. By comparing (b) and (c) in Fig.~\ref{fig:Quantization-phase-shift}, we can see that NNCTC can more accurately fit the waveform, and its amplitude values are more precise compared to WEBee. Comparing (e) and (f) in Fig.~\ref{fig:Quantization-phase-shift} shows the signal phase differences between NNCTC and WEBee.

\subsection{NNCTC Based on CCK}
\begin{figure}[!tbp]
    \centering
    \subfloat[PRR vs. Distance]{\includegraphics[width=0.5\linewidth]{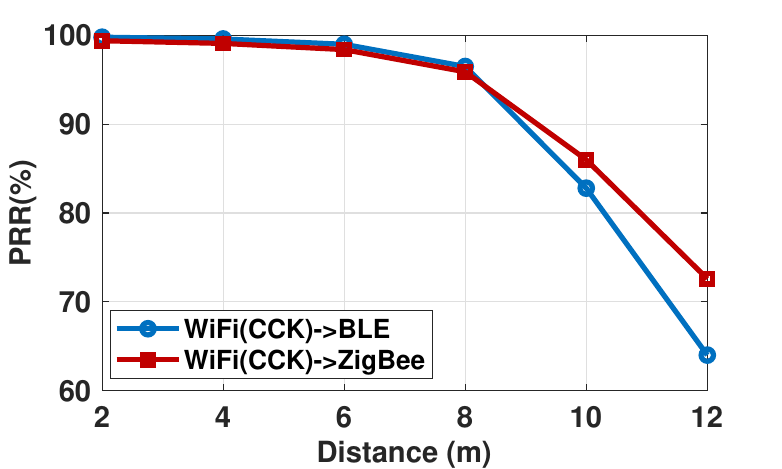}}
    \subfloat[PRR vs. Tx Power]{\includegraphics[width=0.5\linewidth]{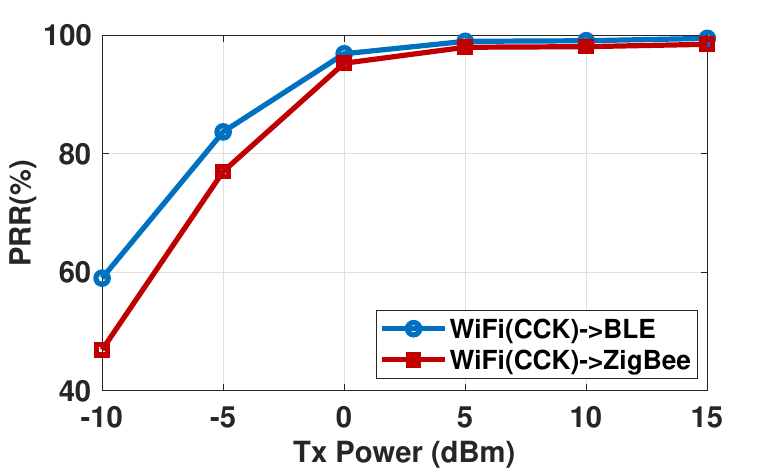}} 
    \caption{Evaluation of Packet Reception Ratio (PRR) for CTC emulation from CCK to BLE and ZigBee using NNCTC methodology.}
    \label{fig:cck_eavl_prr}
    \vspace{-2mm}
\end{figure}

NNCTC is a method model that utilizes interpretable and lightweight neural networks to achieve traditional CTC. In addition to the OFDM-based CTC mentioned earlier, we have also designed and constructed lightweight neural network models based on the NNCTC concept to implement CCK to BLE and CCK to ZigBee. It is worth noting that while both were implementing CTC from WiFi to BLE/ZigBee, OFDM is more challenging than CCK, which is one reason we chose OFDM as a case study. CCK modulation lacks the CP errors inherent in OFDM and exhibits signals closer to those of BLE/ZigBee. As shown in Fig.~\ref{fig:cck_eavl_prr}, we evaluated the PRR capability of NNCTC based on CCK. Experiments were conducted indoors using USRP B210 as the transmitter and CC2650 as the receiver. Fig.~\ref{fig:cck_eavl_prr}(a) illustrates PRR at different transmission distances measured at a fixed transmission power of 10dBm, while Fig.~\ref{fig:cck_eavl_prr}(b) depicts PRR corresponding to different transmission powers at a fixed distance of 2.5m. According to the experimental results, it is observed that when the distance is less than or equal to 6m, or the transmission power is greater than or equal to 5dBm, NNCTC achieves a high Packet Reception Ratio (PRR) for the implementation of CCK to ZigBee and BLE, exceeding 99.0\% at its peak. This is comparable to the performance of classic CCK-based CTC implementations such as WiBeacon\cite{liu2021wibeacon} and WibZig\cite{cheng2023wibzig}, both achieving optimal PRR exceeding 99\% and supporting commercial devices.

\subsection{Training Costs and Comparisons}
\begin{table}[!tbp]
  \caption{Typical Training Costs of NNCTC and Comparison with XiTuXi}
  \label{tab:cost_eval}
  \begin{tabular}{ccl}
    \toprule
    Items&XiTuXi*&NNCTC\\
    \midrule
    Training Time & \shortstack{$>90\,h$} & $\leq10\,min$\\
    Inference Time & $\approx1\,s$ & $\approx3\,min$ \\
    Data Scale & $32K$ & $0.1K$\\
    Corpus Dependency & Dependency & Independent\\
    Network Complexity & High & Low \\
    Interpretability & Low & High\\
  \bottomrule
\end{tabular}\\
*Note: The setup of XiTuXi is based on \cite{liao2023xituxi}.
\vspace{-3mm}
\end{table}
As shown in Table~\ref{tab:cost_eval}, we compared several common training settings between NNCTC and XiTuXi. 
In terms of training and inference time, due to the utilization of lightweight models by NNCTC, completion of NN model training can be achieved in a relatively short period (with a dataset size of 0.1K). For NNCTC's inference time, we encompass the entire process, from ZigBee payload to WiFi payload, with the primary time consumption occurring during channel encoding emulation, requiring approximately 3 minutes per inference. Conversely, according to \cite{liao2023xituxi}, XiTuXi necessitates training for over a hundred hours before usability is attained. Furthermore, NNCTC only requires a small dataset, and once the parameters are assigned to the model, training may not be necessary. Additionally, NNCTC has greater advantages in terms of corpus dependency, model complexity, and interoperability. Regarding training setup, we utilized MATLAB's ZigBee/BLE toolbox to generate standard waveforms as training data. Subsequently, we employed the Adam optimizer and placed the model on a CPU with Intel i9-12900, coupled with 64GB RAM, and an RTX3060 GPU with 12GB of memory.

\section{DISCUSSION AND LIMITATIONS}\label{sec: DISCUSSION}
\textbf{Generalized CTC Models.}
When dealing with different wireless technologies, achieving better performance often requires training with corresponding datasets because different wireless signals exhibit distinct signal characteristics(Such as, between BLE and LoRa). Designing a universal, non-repetitive training model to implement CTC still has a long way to go, and we will consider this in future research.

\textbf{Application Scenarios.}
One potential application scenario for NNCTC and traditional CTC is in assisting emergency communications. For instance, in the event of a wildfire where a ZigBee network gateway is damaged, we can swiftly control a WiFi drone to temporarily act as a ZigBee-WiFi gateway. Furthermore, another advantage of NNCTC is that its neural network design can better leverage the acceleration provided by GPUs, thus working more efficiently.

\textbf{The necessity of on-board NN inference.}
The capability of on-board NN inference can help devices better fulfill CTC tasks. Firstly, it can reduce the bandwidth consumption associated with transferring CTC mapping tables to remote servers, enabling offline processing capabilities. Secondly, achieving CTC mappings between different technologies by exhaustively considering all possible mapping tables is both complex and unnecessary, making on-board NN online inference more efficient and saving local storage space. On-board NN inference also allows IoT devices to dynamically compute optimal CTC mapping tables based on environmental conditions and requirements. Therefore, the inference capability of on-board NNs will be crucial in the future.

\textbf{Online Real-time Processing of NNCTC.}
The lightweight neural network model lays the foundation for online processing, but there still exists a certain time consumption in the CTC calculation process. One potential solution is to pre-store all possible mapping tables locally, which can significantly reduce the time required for CTC payload conversion. In the future, we will continue to explore this approach.

\section{Related Work}\label{sec: related_work}
\textbf{Traditional CTC.}
Traditional CTC work can be broadly categorized into two classes: one represented by packet-level CTC, with FreeBee\cite{kim2015freebee} as a classic example, and the other by physical-layer CTC, with WEBee\cite{li2017webee} as a representative. Packet-level CTC is conceptually simpler and typically utilizes signal features, such as \cite{guo2020zigfi, zheng2018stripcomm, jiang2018achieving, yin2017c, chebrolu2012esense, wang2018symbol, xia2021c, wang2023ligbee}. However, its major drawback is unreliability and low performance. \cite{li2017webee} first introduced a strategy for implementing CTC at the physical layer through signal simulation.
The purpose of the emulation is to select the appropriate WiFi data payload to transmit the WiFi data frame containing the CTC signal we want (related work such as ZigBee, LoRa, BLE, respectively in \cite{li2017webee,xia2022wira, li2021wible}). CTC implemented through simulation strategies often uses WiFi as the transmitting end because WiFi has higher computational capabilities and is generally directly powered. Since WiFi has numerous versions, corresponding CTC work based on WiFi has exhibited some differences, such as work based on IEEE 802.11b (CCK)\cite{cheng2023wibzig, liu2021wibeacon, li2019physical, gawlowicz2022wi}, work based on IEEE 802.11a/g/n/ac (OFDM)\cite{xia2022wira, li2021wible, guo2019wide}, and even work based on IEEE 802.11ax (OFDMA)\cite{xia2023parallel}. In addition to simulation strategies, \cite{liu2020xfi} has also introduced the concept of signal piggybacking to achieve CTC for low-rate IoT devices to WiFi. In addition, physical-layer CTC work also includes \cite{li2018longbee, chen2018twinbee, li2020bluefi, shi2019lorabee, li2020ble2lora, liu2019lte2b}. Traditional CTC relies on empiricism and requires skilled professionals to carefully configure parameters.

\textbf{Neural Network-Based Physical Layer.}
In wireless communication protocols, the physical layer is the most fundamental and crucial layer, often responsible for signal modulation and demodulation. Recently, some research efforts have attempted to construct the physical layer using neural networks. For instance, \cite{wang2023demo}, following a model-driven approach, developed a neural network-based physical layer modulator. The work \cite{soltani2022neural} constructed an OFDM receiver based on neural networks and deployed it on resource-constrained IoT devices. Experiments showed that the modular receiver based on neural networks exhibited improved bit error rate performance. The work \cite{zhang2020deepwiphy} introduced a deep learning framework called DeepWiPHY, which aims to replace certain functionalities in IEEE 802.11ax. Through data-driven training of neural networks, it can achieve WLAN receivers with performance equal to or surpassing traditional networks. The work \cite{zhao2021deep} developed a deep complex convolutional network (DCCN) to replace the FFT processor in OFDM receivers. In addition, the work \cite{o2017introduction, soltani2018autoencoder, gao2018comnet, mu2019end} demonstrated how to integrate neural networks into the physical layer. 
In addition, some work has focused on transforming certain DSP operations into neural networks, such as Sionna\cite{hoydis2022sionna} and DDSP\cite{engel2020ddsp}. Our main contribution lies in how to extend these ideas to wireless environments and improve the efficiency of several key components by using native neural network structures, such as converting DFT into transposed convolutional layers.

\textbf{Large-Model-Based CTC.}
We have noticed that recent research endeavors have attempted to implement CTC using large models, such as \cite{liao2023xituxi} utilizing Neural Machine Translation (NMT) for automatic CTC. There are fundamental differences between this work and our approach. Firstly, complex network models are suitable for running on cloud servers but not well-suited for direct deployment on real-time devices. In contrast, the NNCTC proposed in this paper aims to establish lightweight network models for seamless integration with current NN-based PHY-Layer solutions. Secondly, the training process for large models demands vast training datasets, indicating that CTC based on large models still requires traditional CTC to generate rich datasets. Essentially, it can be viewed as an application or service at the top layer of CTC, but cannot fully replace CTC. Therefore, we argue that our NNCTC in this paper still contributes significantly and can complement CTC based on large models.


First and foremost, traditional CTC strategies have laid a robust foundation for the development of subsequent CTC methods. This paper is based on the framework of traditional CTC and aims to enhance CTC intelligently by introducing neural networks. The goal is to simplify complex parameter configurations by incorporating neural networks and seamlessly integrating with the latest NN-based PHY-Layer solutions. Additionally, traditional CTC relies on empiricism, and by harnessing the powerful learning capabilities of neural networks, it is possible to discover improved parameters, thereby enhancing CTC performance.

\section{Conclusion}\label{sec: conclusion}
This paper shifts the focus of traditional CTC implementation strategies towards neural networks. By embracing this idea, we can establish neural networks to automatically implement CTC across various network protocols. Furthermore, neural networks can optimize traditional CTC approaches, flexibly seeking the optimal solutions. In summary, leveraging neural networks propels CTC into a new era of intelligence, offering possibilities for future native AI in the PHY-Layer.

\begin{acks}
This research/project is supported by the Ministry of Education, Singapore, under its Academic Research Fund Tier 2 (MOE-T2EP20221-0017). This research/project is supported by the National Research Foundation, Singapore and Infocomm Media Development 
Authority under its Future Communications Research \& Development Programme. This work was supported by The Future Network Scientific
Research Fund Project (Grant No. FNSRFP-2021-YB-17).
\end{acks}

\bibliographystyle{ACM-Reference-Format}
\bibliography{ref}
\end{document}